\documentstyle[12pt]{article}

\begin{document}

\title{The MSW effect in a fluctuating matter density}
\author{A.~B.~Balantekin\thanks{E-mail address: {\tt
		baha@nucth.physics.wisc.edu}}
	\ and J.~M.~Fetter\thanks{E-mail address: {\tt
		fetter@nucth.physics.wisc.edu}}\\
	Department of Physics, University of Wisconsin at Madison\\
	Madison, WI 53706
      \and
	F.~N.~Loreti\thanks{Present address:
		Systems for Market Research,
		Pittsburgh, PA 15229; E-mail address: {\tt
		FNLoreti@aol.com}}\\
	Institute for Nuclear Theory, University of Washington\\
	Seattle, WA 98195}
\date{\today}
\maketitle

\begin{abstract}
We consider the effect on matter-enhanced neutrino flavor
transformation of a randomly fluctuating, delta-correlated matter
density.  The fluctuations will produce a distribution of neutrino
survival probabilities.  We find the mean and variance of the
distribution for the case of solar neutrinos, and discuss the
possibility of placing a limit on solar density fluctuations using
neutrino data.
\end{abstract}

\baselineskip=24pt

\section{Introduction}\label{sec:intro}

Matter-enhanced neutrino oscillations, especially in connection to
the solar neutrino problem \cite{baha}, have been extensively studied
in the recent years. More recently some interest has developed in the
problem of neutrino flavor transformations via the
Mikheyev-Smirnov-Wolfenstein (MSW) effect in a randomly fluctuating
matter density.  A general approach to neutrino oscillations in such
inhomogeneous matter was developed in Ref.~\cite{sawyer}.  A
Boltzmann-like collision integral with blocking factors, describing
the decoherence of neutrinos in matter, was given in
Ref.~\cite{raffelt}.  Matter fluctuations which are not random, but
harmonic \cite{koonin, haxton} or changing stepwise \cite{ks-step}
were also considered.

Redfield equations for a neutrino traveling in a region with
delta-correlated Gaussian noise were recently developed in
Ref.~\cite{us1} and applied to two-neutrino flavor transformations in
the post-core bounce supernova environment in Ref.~\cite{us2}. In
parallel to these papers, an analytical procedure to calculate the
survival probability was described in Ref.~\cite{lujan}, and further
implications of solar matter density random noise upon resonant
neutrino conversion were studied in Ref.~\cite{nrsv}.

The aim of this paper is to expand the analysis of Ref.~\cite{us1} to
investigate the mean and variance of the distribution of neutrinos
when a randomly-fluctuating, delta-correlated electron density is
present in the sun.  A general treatment of fluctuations is presented
in Section \ref{sec:fluc}.  Mean survival probabilities and the
variances of the survival probability distribution are given in
Sections \ref{sec:mean} and \ref{sec:higher}, respectively.  In
Section 5, we discuss the results and present conclusions.

\section{General Treatment of Fluctuation}\label{sec:fluc}

We will be concerned with systems whose evolution is described by the
Schr\"odinger-like equation
\begin{equation}
i \frac{\partial}{\partial t} \psi(t) = H(t) \psi(t),
\end{equation}
where $\psi$ is represented as a column vector.  Following the method
of Loreti and Balantekin \cite{us1}, we define the density matrix
\begin{equation}
\rho = \psi \otimes \psi^{\dag}
\end{equation}
and divide H into two parts, one with known time dependence and one
which fluctuates with time:
\begin{equation}
H(t) = H_0(t) + B(t) M,
\end{equation}
where $B(t)$ is a c-number and the operator $M$ does not depend on
time.

We assume that the fluctuation $B(t)$ obeys
\begin{eqnarray}
\langle B(t_1) \rangle & = & 0 \nonumber \\
\langle B(t_1) B(t_2) \rangle & = & \alpha^2 f_{12} \nonumber \\
\langle B(t_1) B(t_2) B(t_3) \rangle & = & 0 \nonumber \\
\langle B(t_1) B(t_2) B(t_3) B(t_4)\rangle & = & \alpha^4
	(f_{12} f_{34} + f_{13} f_{24} + f_{14} f_{23}) \\
& \vdots \nonumber &
\label{eq:fluct-cond}
\end{eqnarray}
where $f_{ij} = f(|t_j - t_i|)$ gives the correlation between
fluctuations in different places.  Throughout the current paper, we
will consider the case of delta-correlated (white) noise:
\begin{equation}
f(x) = 2\tau\delta(x),
\end{equation}
with the correlation time $\tau$ as a parameter.  This is equivalent
to the statement that the probability of a given $B(t')$ is
proportional to
\begin{equation}
\int_0^t \exp \left\{ [B(t')]^2 / 2 \tau \alpha^2 \right\} \,dt'.
\end{equation}

Further, the results from delta correlations will be approximately
the same as those from step-function correlations of step length
$\tau$ as long as we have the constraint
\begin{equation}
\tau \ll \left(C(t) + \frac{d}{dt} \log\alpha(t)\right)^{-1},
\label{eq:step-approx}
\end{equation}
where $C(t)$ is the largest element of the matrix commutator $[H_0,
M]$.  Then we have the following result for the evolution of
fluctuation-averaged $\rho$ (cf.~Eq.~(16) of Ref.~\cite{us1}):
\begin{equation}
\frac{\partial}{\partial t} \langle \rho(t) \rangle = 
-\alpha^2 \tau [M, [M, \langle \rho(t) \rangle]] 
- i[H_0(t), \langle \rho(t) \rangle].
\label{eq:avgev}
\end{equation}

MSW conversion between neutrino flavors obeys the equation
\begin{equation}
i \frac{\partial}{\partial t}
 \left( \begin{array}{c} \nu_e \\ \nu_x \end{array} \right) =
\frac{\Delta m^2}{4 E}
\left( \begin{array}{cc}
\zeta(t) - \cos 2\theta & \sin 2\theta \\
\sin 2\theta            & -(\zeta(t) - \cos 2\theta)
\end{array} \right)
\left( \begin{array}{c} \nu_e \\ \nu_x \end{array} \right)
\label{eq:msw}
\end{equation}
where
\begin{equation}
\zeta(t) = \frac{2 \sqrt{2} G_F E}{\Delta m^2} N_e(t),
\end{equation}
$\theta$ is the vacuum neutrino mixing angle, $\Delta m^2$ is the
difference in the squared masses of the two neutrino species, $E$ is
the neutrino energy, and $N_e$ is the electron number density.

In order to study the influence of density fluctuations on the
conversion rate, one can add periodic matter density perturbations to
the average density \cite{koonin}
\begin{equation}
N_e(r) = \overline{N_e} \, [ 1 + \epsilon \sin (kr) ], 
\end{equation}
where the wavenumber $k$ is fixed. Such an additional term can induce
additional MSW level-crossings giving rise to interference between
them. To elucidate this behavior, one can utilize logarithmic
perturbation theory, valid for small vacuum mixing angles. The
application of logarithmic perturbation theory to the neutrino mixing
problem is sketched in the Appendix. If there is more than one MSW
resonance point, one can calculate the integral in
Eq.~(\ref{eq:psiint}) to obtain the electron neutrino survival
probability, $P_e$, as
\begin{equation}
P_e = \exp \left[ - {\pi \Delta m^2 \sin^ 2 \theta \over 4 E} \left|
\sum_{t_a} [N_e' (t_a)]^{-1/2} 
\exp \left[ i\int_0^{t_a} \frac{\Delta m^2}{2E}
[\zeta(t')-\cos 2 \theta] dt' \right] \right|^2 \right]
\end{equation}
If two turning points are close enough, one can utilize the uniform
Airy approximation to obtain
\begin{equation}
P_e = \exp \left[ - \pi \frac{\Delta m^2 \sin^2 2
         \theta}{2E|\zeta'(t_a)|} \sin^2 \left( \frac{\pi}{4} -
         \frac{\Delta m^2}{4E} \int_{t_a}^{t_b} [\zeta(t) - \cos 2
         \theta] \, dt \right) \right],
\end{equation}
which is the wavenumber dependence observed in Ref. \cite{koonin}.

If the density fluctuations cannot be parameterized with a
single wavenumber as given above, but with a distribution of
wavenumbers, one can then write
\begin{equation}
N_e (r) = \overline{N_e (r)} \, [ 1 + \int dk \, \epsilon (k) \sin
(kr)].
\end{equation}
The periodic fluctuation with a given wavenumber $k_0$ can be
recovered by setting $\epsilon (k) = \epsilon \delta (k-k_0)$. The
density fluctuations given in Eqs.~(\ref{eq:fluct-cond}) can then be
considered as resulting from stochastically distributed
$\epsilon(k)$'s. 

\section{Calculation of the Mean Survival Probability}\label{sec:mean}

We will assume that the electron density $N_e$ fluctuates around the
value $\overline{N_e}$, given by the Bahcall-Pinsonneault Standard
Solar Model (SSM) including helium diffusion:
\begin{equation}
N_e = (1 + \beta) \overline{N_e}
\end{equation}
where $\beta$ fluctuates and obeys constraints similar to those of
Eq.~(\ref{eq:fluct-cond}) with delta correlations.

We may then use the formalism of Section \ref{sec:fluc} with
\begin{equation}
H_0 = \sigma_z A(t) + \sigma_x B,
\quad
M = \sigma_z,
\end{equation}
where we have defined
\begin{equation}
A(t) \equiv \frac{\Delta m^2}{4E} (\overline{\zeta(t)} - \cos
2\theta), \quad B \equiv \frac{\Delta m^2}{4E} \sin 2\theta.
\end{equation}
Then \cite{us1} we have the result
\begin{equation}
\frac{\partial}{\partial t}
\left( \begin{array}{c} r \\ x \\ y \end{array} \right)
= -2 \left( \begin{array}{ccc}
 0  &     0  &      B \\
 0  &     k  &  -A(t) \\
-B  &  A(t)  &      k
\end{array} \right)
\left( \begin{array}{c} r \\ x \\ y \end{array} \right)
\label{eq:meanev}
\end{equation}
where
\begin{equation}
k \equiv 2 \langle \beta^2 \rangle \tau,
\end{equation}
and
\begin{eqnarray}
r &=& 2 \langle \nu_e^* \nu_e^{\phantom{*}} \rangle - 1 \nonumber \\
x &=& 2 \> \hbox{Re} \, \langle \nu_\mu^* \nu_e^{\phantom{*}} \rangle
	\nonumber \\
y &=& 2 \> \hbox{Im} \, \langle \nu_\mu^* \nu_e^{\phantom{*}} 
	\rangle.
\end{eqnarray}

The condition of Eq.~(\ref{eq:step-approx}) now becomes
\begin{equation}
\tau \ll \left( \sin 2\theta \frac{\Delta m^2}{2E} \right)^{-1}
\label{eq:step-approx-msw}
\end{equation}
where we have assumed that $\log N_e$ varies slowly with time.  The
right-hand side is the oscillation length of neutrinos at the
resonance, divided by $4\pi$.  This condition is similar to that of
Ref.~\cite{nrsv} that $\tau \ll \lambda_m$, the oscillation length of
neutrinos at any given point.  For definiteness, we will take the
correlation length $\tau$ to be 10 km.  Then the constraint
(\ref{eq:step-approx}) becomes (assuming that the logarithmic
derivative is small, which is accurate for the sun)
\begin{equation}
\sin 2\theta \> \Delta m^2 / E \ll 3.95 \times 10^{-5} \hbox{ eV}^2/\hbox{MeV}.
\label{eq:tau-constraint}
\end{equation}
We will present some results for which this condition does not hold,
that is, for which the delta correlations we assume would give a
different result from step-function correlations.  The advantage of a
constant correlation length is that the parameter $k$ is constant in
$\Delta m^2 / E$, so that we can more meaningfully compare the
effects of fluctuations with different MSW parameters than if we let
$\tau$ vary to satisfy Eq.~(\ref{eq:step-approx}).  In any case,
since $\tau$ and $\langle \beta \rangle_{\hbox{\scriptsize rms}}$
enter only through $k$, one can get identical results to ours by
making $\tau$ smaller and increasing $\langle \beta
\rangle_{\hbox{\scriptsize rms}}$.

We solved the system (\ref{eq:meanev}) numerically; results are
presented in Figures 1--3 for $\sin^2 2\theta$ equal to 0.01, 0.1,
and 0.7, respectively, and a several values of $\langle \beta
\rangle_{\hbox{\scriptsize rms}}$.  0.01 and 0.7 correspond roughly
to the small- and large-mixing solutions \cite{bhkl} to the solar
neutrino problem.  We used the standard solar model of Bahcall and
Pinsonneault, including helium diffusion \cite{bpssm}, but assumed
all neutrinos were produced at the center of the sun.  Usually,
fluctuations suppress the $\nu_e \to \nu_\mu$ transition.  For the
large-angle solution ($\sin^2 2\theta = 0.7$), the effect on the mean
survival probability is noticeable only in the adiabatic region, even
in the physically unreasonable case $\langle \beta
\rangle_{\hbox{\scriptsize rms}} = 0.5$.  For smaller angles, there
is some effect in the non-adiabatic region, but the greatest effect
is still in the adiabatic region.

In the adiabatic region, for survival probabilities greater than
$1/2$, fluctuations tend to enhance the transition.  This has been
explained \cite{us1} by appealing to the flatness of the
Bahcall-Pinsonnealt density profile near the center.  Neutrinos in
this region do not go through the resonance point, but do go through
part of the resonance region.  Travel through that region has a
significant effect since the density profile is flat just after
production at the center.  Fluctuations can therefore have an effect.
To confirm that the enhancement depends on the flatness of the
density profile, we have also solved the system of equations
(\ref{eq:meanev}) for an exponential density profile with central
density $6.25 \times 10^{25}$ cm$^{-3}$, and scale height $6.21
\times 10^{9}$ cm.  These values were chosen to give the same density
as the Bahcall-Pinsonneault model at the center and edge of the sun.
At $r=R_\odot$, we cut the density to zero for both profiles.  The
exponential results are compared to the Bahcall-Pinsonneault results
in Figures 4--6.  For fluctuations above about 2\%, the exponential
profile gives noticeably weaker enhancement of the transition
probability than the Bahcall-Pinsonneault profile, and changes the
shape of the curve elsewhere as well.  This result indicates that
large fluctuations can induce appreciable non-adiabatic effects in
the adiabatic region, since the initial density is the same in both
cases, and both have the same numerical value of final density before
the truncation to zero.  (In any case, the step should have no
effect, since the density at the edge of the sun is much less than
the resonant density.)

\section{Higher Moments}\label{sec:higher}

The formalism developed in Ref.~\cite{us1} may be used to calculate
higher moments of the distribution of survival probabilities.  We
will present results for the variance of the distribution,
\begin{displaymath}
\sigma^2 \equiv \langle P_e^2 \rangle - \langle P_e \rangle^2.
\end{displaymath}
In principle, the formalism could be used to calculate arbitrary
higher moments of the distribution of $P_e$.

From the evolution equation (\ref{eq:msw}), it is straightforward to
show that
\begin{equation}
i \frac{\partial}{\partial t}
    \left( \begin{array}{c}
	\nu_e^* \nu_e   \\ \nu_\mu^* \nu_\mu \\
	\nu_e \nu_\mu^* \\ \nu_e^* \nu_\mu
    \end{array} \right) =
\frac{\Delta m^2}{4 E}
\left( \begin{array}{cc}
0                           & (\sigma_x - 1) \sin 2\theta \\
(\sigma_x - 1) \sin 2\theta & 2 \sigma_z (\zeta(t) - \cos 2\theta)
\end{array} \right) 
    \left( \begin{array}{c}
	\nu_e^* \nu_e   \\ \nu_\mu^* \nu_\mu \\
	\nu_e \nu_\mu^* \\ \nu_e^* \nu_\mu
    \end{array} \right).
\end{equation}
As before, we define the density matrix $\rho$, which now contains
the square of the neutrino survival probability, and use the
formalism of Section \ref{sec:fluc} with
\begin{equation}
H_0 =
\left( \begin{array}{cc}
0 &
(\sigma_x - 1) B \\
(\sigma_x - 1) B &
2 \sigma_z A
\end{array} \right),
\quad
M =
\left( \begin{array}{cc}
0 & 0 \\ 0 & 2 \sigma_z
\end{array} \right).
\end{equation}

Then in terms of the quantities
\begin{eqnarray}
s &=& \langle \nu_e^* \nu_e^{\phantom{*}} \nu_\mu^* 
	\nu_\mu^{\phantom{*}} \rangle
	= \langle P_e \rangle - \langle P_e^2 \rangle \nonumber \\
q &=& \langle (\nu_e^* \nu_e^{\phantom{*}} 
	- \nu_\mu^* \nu_\mu^{\phantom{*}})
	\nu_\mu^* \nu_e^{\phantom{*}} \rangle \nonumber \\
z &=& \langle (\nu_\mu^* \nu_e^{\phantom{*}})^2 \rangle,
\end{eqnarray}
we have
\begin{equation}
\frac{\partial}{\partial t}
\left( \begin{array}{c} s \\ q_r \\ q_i \\ z_r 
	\\ z_i \end{array} \right)
= -2
\left( \begin{array}{ccccc}
  0 &   0 &  -B &   0 &   0 \\
  0 &   k &  -A &   0 &   B \\
 3B &   A &   k &  -B &   0 \\
  0 &   0 &   B &  4k & -2A \\
  0 &  -B &   0 &  2A &  4k
\end{array} \right)
\left( \begin{array}{c} s \\ q_r \\ q_i \\ z_r \\ z_i \end{array}
\right) + \left( \begin{array}{c} 0 \\ 0 \\ B \\ 0 \\ 0 \end{array}
\right),
\label{eq:varev}
\end{equation}
where the subscripts $r$ and $i$ denote the real and imaginary part
of the quantity.

We again solved the system numerically, assuming fluctuations of the
same form as in Section \ref{sec:mean}.  The results are presented in
Figures 7--9, for the same mixing parameters and values of $\langle
\beta \rangle_{\hbox{\scriptsize rms}}$ as in Figure 1--3.  The
quantity plotted is $\sigma \equiv \sqrt{\sigma^2}$, rather than the
variance itself.  For the smaller angles, there is a dramatic peak in
$\sigma$ around $\Delta m^2 / E = 1.6 \times 10^{-5}$ eV$^2$/MeV,
corresponding to neutrinos being produced in the resonance region
(since we assume that all neutrinos are produced at the center of the
sun).  The results for $\sin^2 2\theta=0.7$ show a similar increase
around $\Delta m^2 / E = 1.6 \times 10^{-6}$ eV$^2$/MeV, which again
corresponds to production inside the resonance region.  It does not
go back down, since the resonance region will include the core for
any $\Delta m^2 / E$ larger than that.

As the rms noise amplitude increases, the variance seems to saturate
at a value of $1/12$, independent of the mixing parameters.  Two
points are relevant to this phenomenon.  First, a uniform probability
distribution bounded between zero and one has the same mean and
variance.  Second, fluctuations will tend to suppress the oscillation
of $\langle P_e \rangle$ as a function of time, since it becomes a
superposition of oscillations with differing phases.  Therefore, we
may set the time derivatives on the left-hand sides of
Eqs.~(\ref{eq:meanev}) and (\ref{eq:varev}) to zero.  Assuming that
$k$ goes to zero smoothly, we are led to the trivial solution $r=0$
for Eq.~(\ref{eq:meanev}), and $s=1/6$ for Eq.~(\ref{eq:varev}),
where we have taken the limit $k=0$.  This yields a variance of
$1/12$ and a mean of $1/2$, as expected.

In order to illustrate the magnitude of the effect, in Figures 10-12,
we plot the mean survival probability plus and minus the width
$\sigma$ of the distribution.  Note that, even though we did not
calculate the higher moments, the mean survival probability plus and
minus $\sigma$ is bounded between zero and one, suggesting that the
distribution of $P_e$ is not very skewed.

\section{Discussion}\label{sec:disc}

The variance is a potentially important tool for the exploration of
solar density fluctuations on some time scales.  Fluctuations on the
time scale of a radiochemical experiment's run would broaden the
distribution of count rates, since different runs would have
different survival probabilities.  None of the experiments currently
operating has noted a broader distribution of rates than expected
\cite{sage, gallex, homestake}, suggesting that the neutrino data
will probably limit, rather than measure, such fluctuations.  For the
favored, small-angle solution, the variance is strongly peaked when
neutrinos are produced near the resonance.  This suggests that the
finite radial distribution of neutrino production will have an
important effect, likely extending the peak to lower values of
$\Delta m^2 / E$, since those correspond to resonance farther out in
the sun.  Further, it means that density fluctuations will affect pp
neutrinos more strongly than other neutrinos, so that gallium
experiments may put the strongest limit on fluctuations.

To develop a very rough estimate of the limit on fluctuations, we
note that their effect should become noticeable when the ratio
$\sigma / \langle P_e \rangle$ becomes comparable to the relative
$1\sigma$ experimental uncertainty.  To simplify the argument, we
assume that the signal observed at the gallium experiment consists
only of $\sim 0.3$ MeV pp neutrinos.  (Standard MSW analyses indicate
the near-complete suppression of other neutrinos.)  As an example, we
will consider the GALLEX results.  GALLEX has a $1\sigma$ uncertainty
of approximately 13\% and an energy-averaged survival probability of
about 60\% \cite{gallex}.  A 1\% density fluctuation on the time
scale of a GALLEX run, 20 to 28 days, with $\sin^2 2\theta=0.01$ and
$\Delta m^2 / E = 1.63 \times 10^{-5}$ eV$^2$/MeV (chosen to give a
60\% survival probability), has $\sigma/\langle P_e \rangle = 0.15$.
It is likely, then, that a careful study could rule out fluctuations
on that level.

In the future, real-time, high-statistics detectors such as HELLAZ
\cite{hellaz} and Borexino \cite{borexino} could be used to
investigate fluctuations on shorter time scales.  In particular, a
helioseismological g-mode oscillation could leave a signature in the
neutrino data.  Unfortunately, the current high-statistics
experiments SNO \cite{sno} and Super-Kamiokande \cite{superk}
probably cannot be used to probe fluctuations in this way, as they
are not sensitive to low-energy neutrinos.

A proper averaging over neutrino production location would be
extremely cpu-intensive.  In this regard, approximate analytic
techniques to compute the moments of the density matrix, such as that
of Ref.~\cite{lujan}, may be very useful in analyzing data.

\section*{Acknowledgments}
We thank John Beacom, George Fuller, and Y.-Z.~Qian for useful
conversations.  This research was supported in part by the
U.S. National Science Foundation Grant No. PHY-9314131 at the
University of Wisconsin, in part the Department of Energy under Grant
No. DE-FG06-90ER40561 at the Institute for Nuclear Theory, and in
part by the University of Wisconsin Research Committee with funds
granted by the Wisconsin Alumni Research Foundation.

\section*{Appendix}

Here we summarize the application of logarithmic perturbation theory
to a two level system \cite{oldbaha}. Consider
\begin{equation}
i \frac{\partial}{\partial t}
  \left( \begin{array}{c} \psi_1 \\ \psi_2 \end{array} \right)
= \left( \begin{array}{cc} A & gC \\ gC & A \end{array} \right)
  \left( \begin{array}{c} \psi_1 \\ \psi_2 \end{array} \right)
\end{equation}
subject to the initial condition $\psi_1(t=0)=1$ and $\psi_2(t=0)=0$.
Defining
\begin{equation}
z \equiv \frac{\psi_2}{\psi_1}
\end{equation}
we find that the quantity $z$ satisfies the Riccati equation,
\begin{equation}
i\dot z = -2Az + gC(1-z^2).
\label{eq:riccati}
\end{equation}
Expanding $z$ in powers of $g$
\begin{equation}
z = z_0 + gz_1 + g^2z_2 + \ldots\ ,
\end{equation}
inserting into Eq.~(\ref{eq:riccati}), and equating powers of $g$, we
find that
\begin{equation}
z = -ig e^{+2i\int_0^t A(t')\,dt'} \int_0^t dt' \, C(t') \; 
  e^{-2i\int_0^{t'} A(t')\,dt''} + {\cal O}(g^3)
\label{eq:powers}
\end{equation}
is the solution with the given initial conditions. In order to
calculate the survival probability, $|\psi_1|^2$, we observe that
\begin{equation}
|\psi_1|^2 = \frac{1}{1 + |z|^2},
\label{eq:psiz}
\end{equation}
where we have used $|\psi_1|^2+|\psi_2|^2=1$. It is easy to show that
Eq.~(\ref{eq:riccati}) can be rewritten as
\begin{equation}
i \frac{d}{dt} \log \left(1 + \left| z \right|^2\right) = gC(z^*-z).
\label{eq:rictwo}
\end{equation}
Eqs.~(\ref{eq:psiz}) and (\ref{eq:rictwo}) then yield
\begin{equation}
|\psi_1|^2 = \exp \left[ 2g\int_0^T dt \, C(t) \, {\rm Im}\, z
\right].
\label{eq:psiinter}
\end{equation}
Finally, substituting the approximate solution,
Eq.~(\ref{eq:powers}), into Eq.~(\ref{eq:psiinter}), we arrive at the
desired answer:
\begin{equation}
|\psi_1|^2 
= \exp \left[ -\frac{g^2}{2} \left| \int_0^T dt \; C(t) \;
  e^{+2i\int_0^t A(t')\,dt'} \right|^2 
  + {\cal O} (g^4) \right].
\label{eq:psiint}
\end{equation}

It usually is possible to calculate the integral in
Eq.~(\ref{eq:psiint}) within the stationary phase approximation. If
the quantities $A(t)$ and $C(t)$ are monotonically changing, there is
only one stationary point $t_a$ and we can approximate the integral
above as
\begin{equation}
|\psi_1|^2 
= \exp \left[ - \pi g^2 \frac{C^2(t_a)}{A'(t_a)} 
  + {\cal O} (g^4) \right],
\end{equation}
where the prime denotes the derivative with respect to time. In most
cases, especially when $A(t)$ fluctuates, there could be more than
one turning point.  If two of the turning points are close to each
other it would be necessary to employ the Airy uniform approximation
\cite{airy}. In this case one gets
\begin{eqnarray}
|\psi_1|^2 & = & \exp \left[ { - \pi^2 g^2 \over 2} 
  \left| \left( \frac{C(t_a)}{\sqrt{|A'(t_a)|}} 
                + \frac{C(t_b)}{\sqrt{|A'(t_b)|}} \right)
         \zeta^{1/4} {\rm Ai}(-\zeta) \right. \right. \nonumber \\
   & & \left. \left. - i \left( \frac{C(t_a)}{\sqrt{|A'(t_a)|}} 
                 - \frac{C(t_b)}{\sqrt{|A'(t_b)|}}\right)
         \zeta^{ -1/4} {\rm Ai}'(-\zeta) \right|^2 
  + {\cal O}(g^4)\right],
\label{eq:airy}
\end{eqnarray}
where $t_a$ and $t_b$ are the turning points, Ai$(x)$ and Ai$'(x)$
are the Airy function and its derivative, and
\begin{equation}
\zeta = \left[ - \frac{3}{2} \int_{t_a}^{t_b} A(t) \, dt
\right]^{2/3}.
\end{equation}
If the turning points are far apart, one can simplify
Eq.~(\ref{eq:airy}) using the asymptotic expressions of the Airy
function:
\begin{equation}
|\psi_1|^2 
= \exp \left[ - 2 \pi g^2 \frac{C^2(t_a)}{|A'(t_a)|}
         \sin^2 \left( \frac{\pi}{4} - \int_{t_a}^{t_b} A(t) \, dt 
	\right) \right],
\end{equation}
where we have assumed that the functions $C(t)$ and $A'(t)$ are very
slowly changing in this interval, i.e.~$C(t_a) = C(t_b)$ and $A'(t_a)
= A'(t_b)$.


\newpage
\section*{Figure Captions}

\noindent
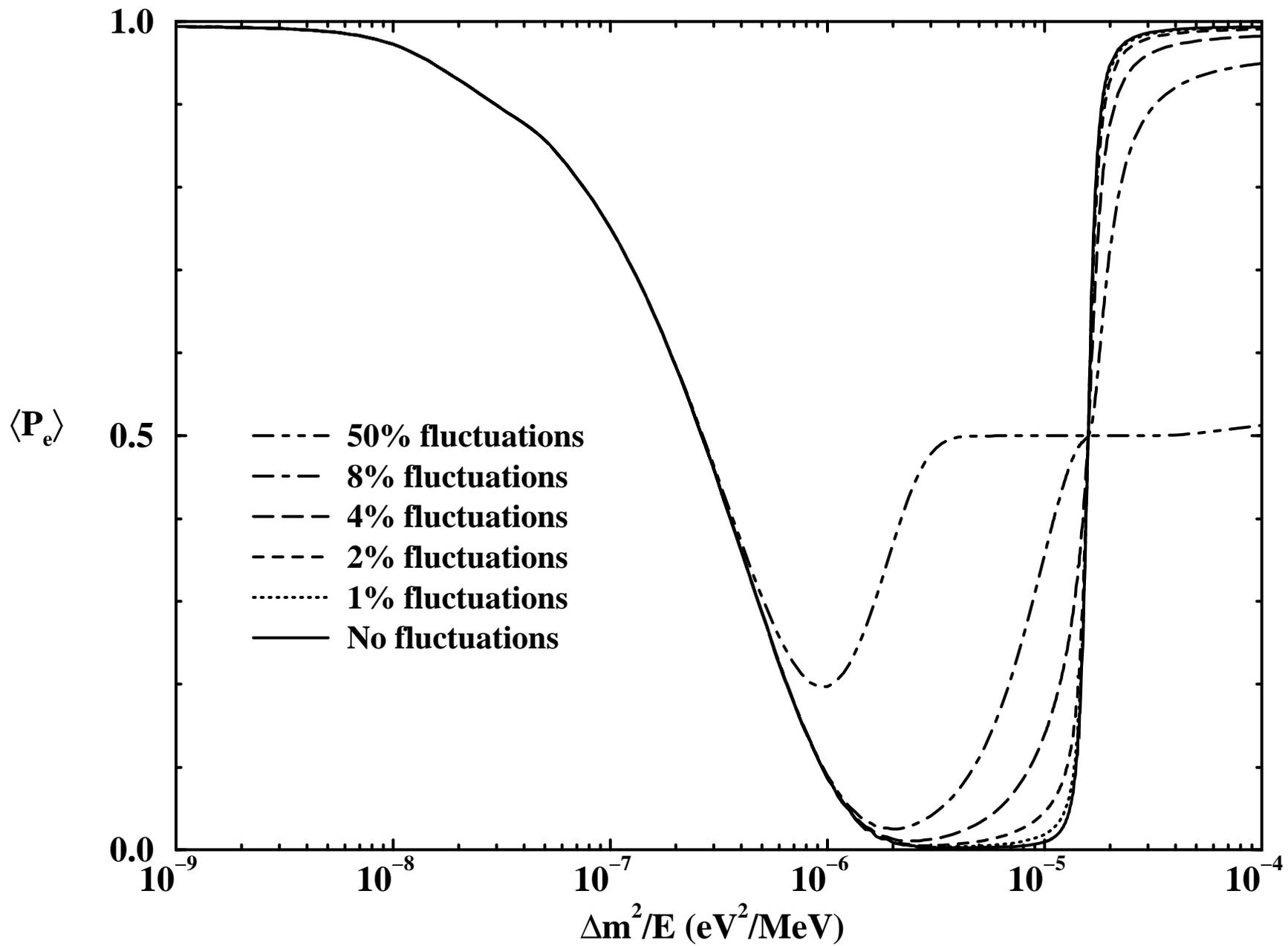
{\bf Figure 1.} Mean survival probability for a Bahcall-Pinsonneault
density profile and $\sin^2 2\theta = 0.01$. The probability is
plotted for $\langle \beta \rangle_{\hbox{\scriptsize rms}}$ equal to
0, 0.01, 0.02, 0.04, 0.08, and 0.5, with a constant correlation
length $\tau=10$ km.
\bigskip

\noindent
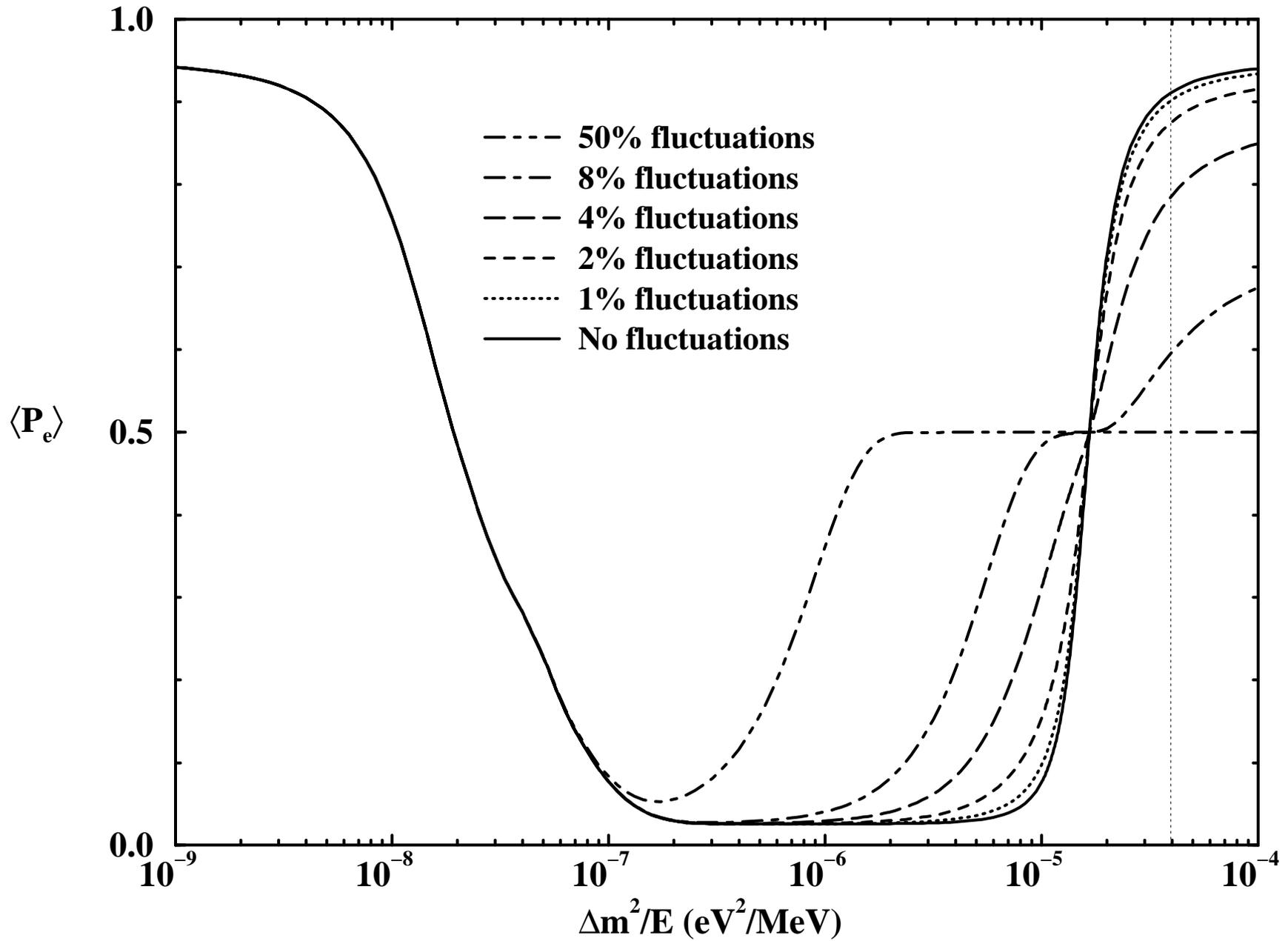
{\bf Figure 2.} As Figure 1, but with $\sin^2 2\theta = 0.1$.  The
vertical dotted line indicates where the constraint of Equation
(\ref{eq:tau-constraint}) begins to break down; there, $\sin 2\theta
\> \Delta m^2 / E = 0.1 \times (3.95 \times 10^{-5})$ eV$^2$/MeV.
\bigskip

\noindent
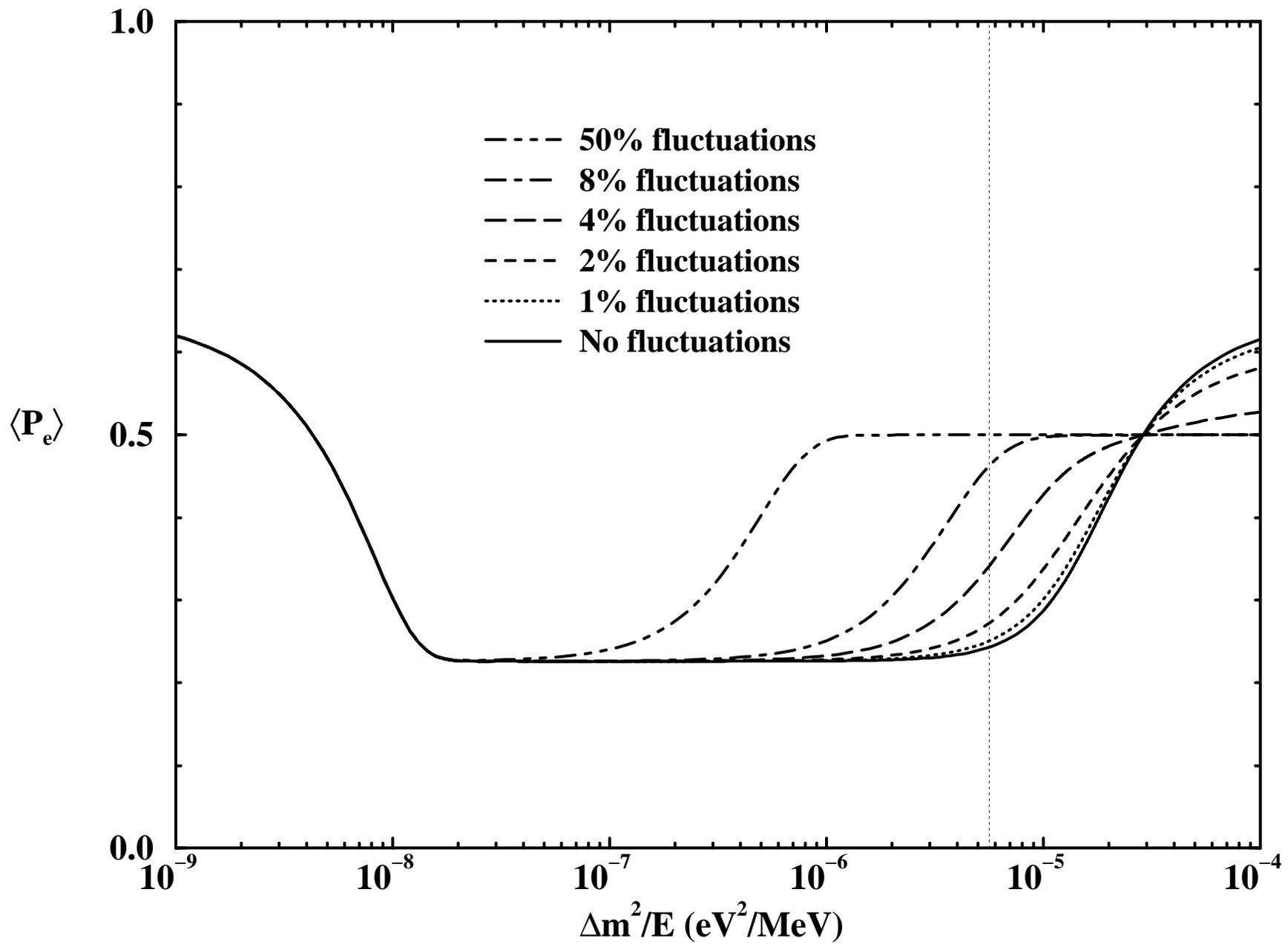
{\bf Figure 3.} As Figure 1, but with $\sin^2 2\theta = 0.7$.  The
vertical dotted line is as in Figure 2.
\bigskip

\noindent
{\bf Figure 4.} Mean survival probability for the
Bahcall-Pinsonneault sun compared to an exponential density profile
with the same central and edge density, with $\sin^2 2\theta = 0.01$.
(a)-(e) show $\langle \beta \rangle_{\hbox{\scriptsize rms}}$ equal
to 0, 0.01, 0.02, 0.04, and 0.08, respectively.  The solid line is
the Bahcall-Pinsonneault sun, and the dashed line is the exponential.
\bigskip

\noindent
{\bf Figure 5.} As Figure 4, but with $\sin^2 2\theta = 0.1$.  The
vertical dotted line is as in Figure 2.
\bigskip

\noindent
{\bf Figure 6.} As Figure 4, but with $\sin^2 2\theta = 0.7$.  The
vertical dotted line is as in Figure 2.
\bigskip

\noindent
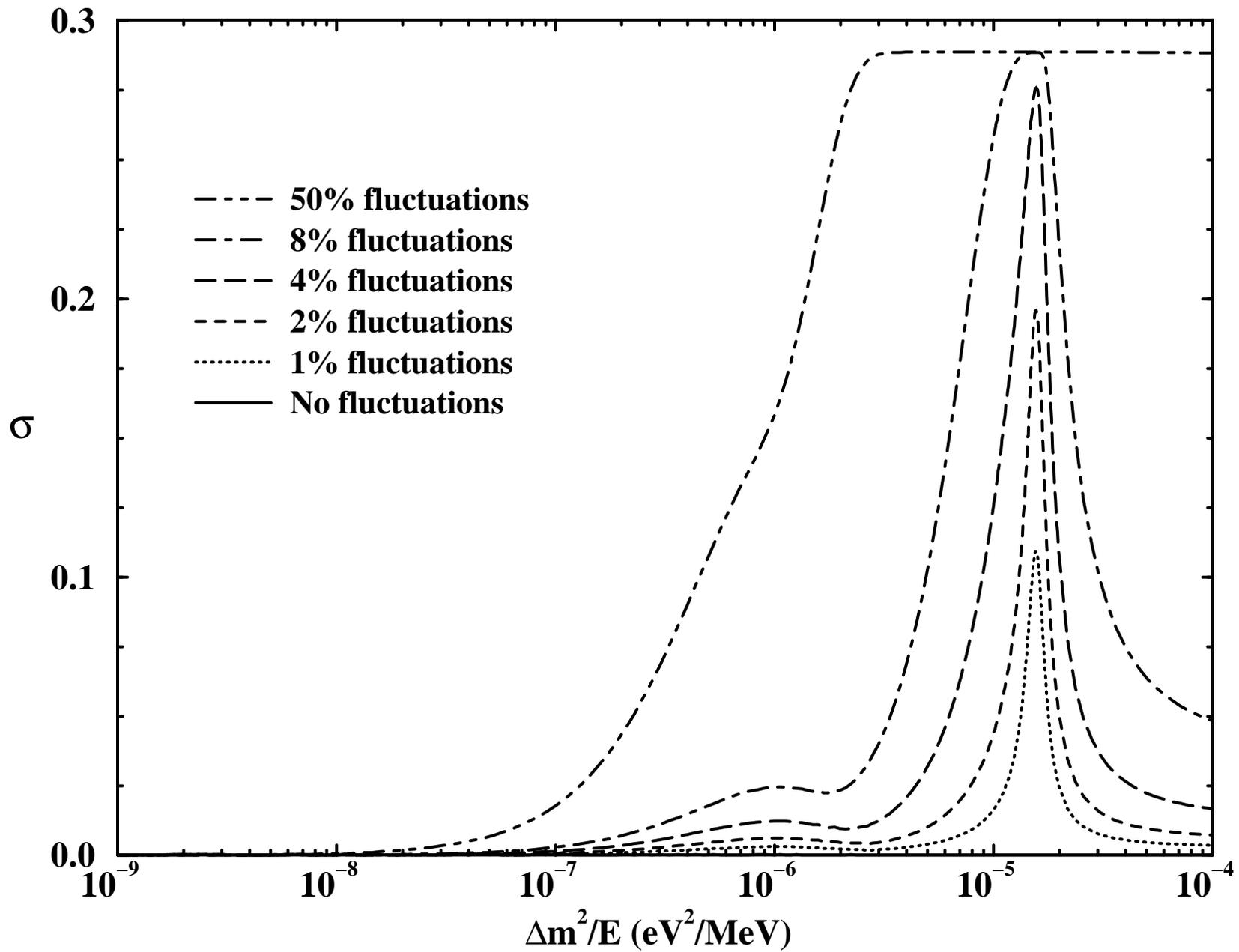
{\bf Figure 7.} $\sigma$ for a Bahcall-Pinsonneault density profile,
with the same values of $\sin^2 2\theta$ and $\langle \beta
\rangle_{\hbox{\scriptsize rms}}$ as Figure 1.
\bigskip

\noindent
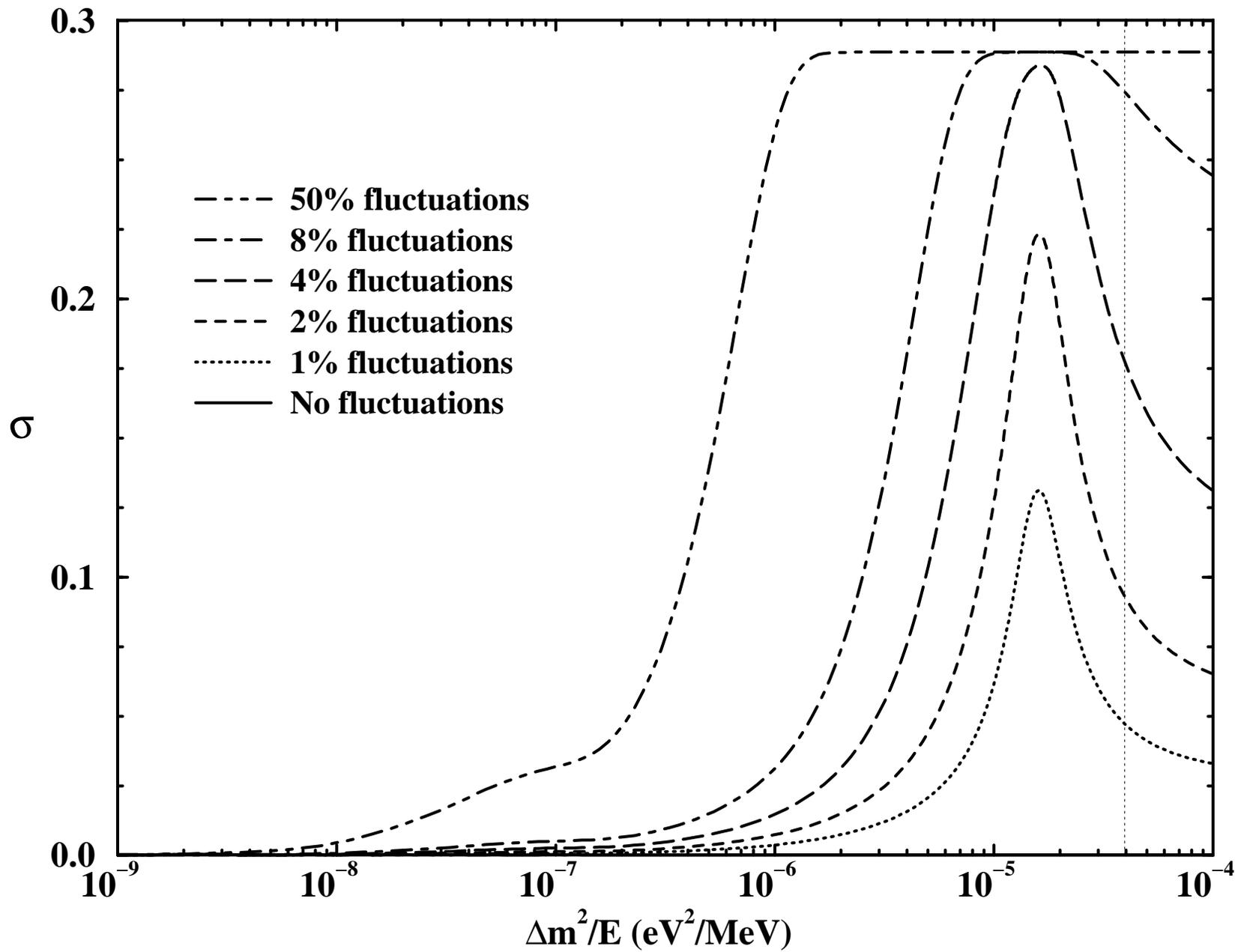
{\bf Figure 8.} As Figure 7, but with $\sin^2 2\theta = 0.1$.  The
vertical dotted line is as in Figure 2.
\bigskip

\noindent
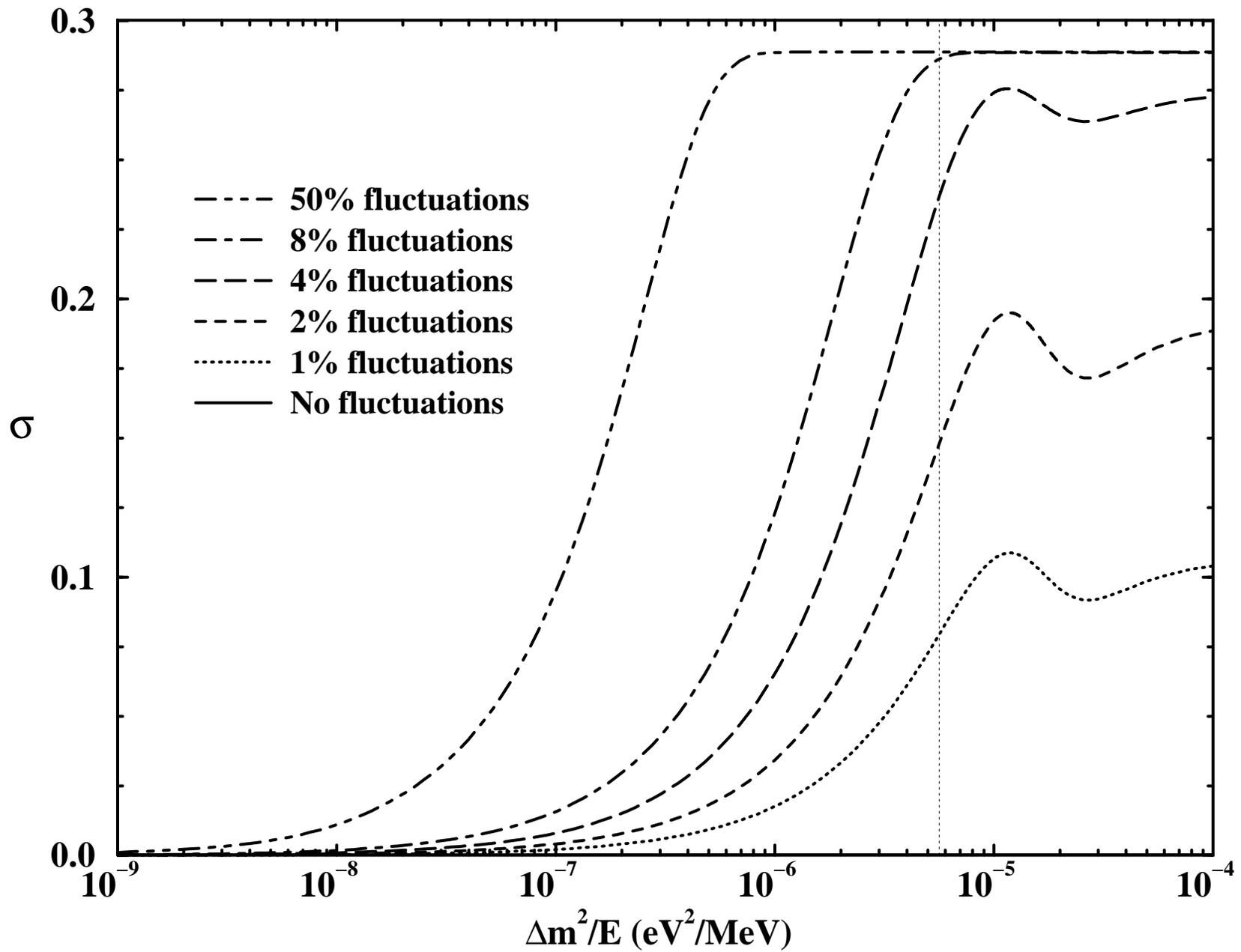
{\bf Figure 9.} As Figure 7, but with $\sin^2 2\theta = 0.7$.  The
vertical dotted line is as in Figure 2.
\bigskip

\noindent
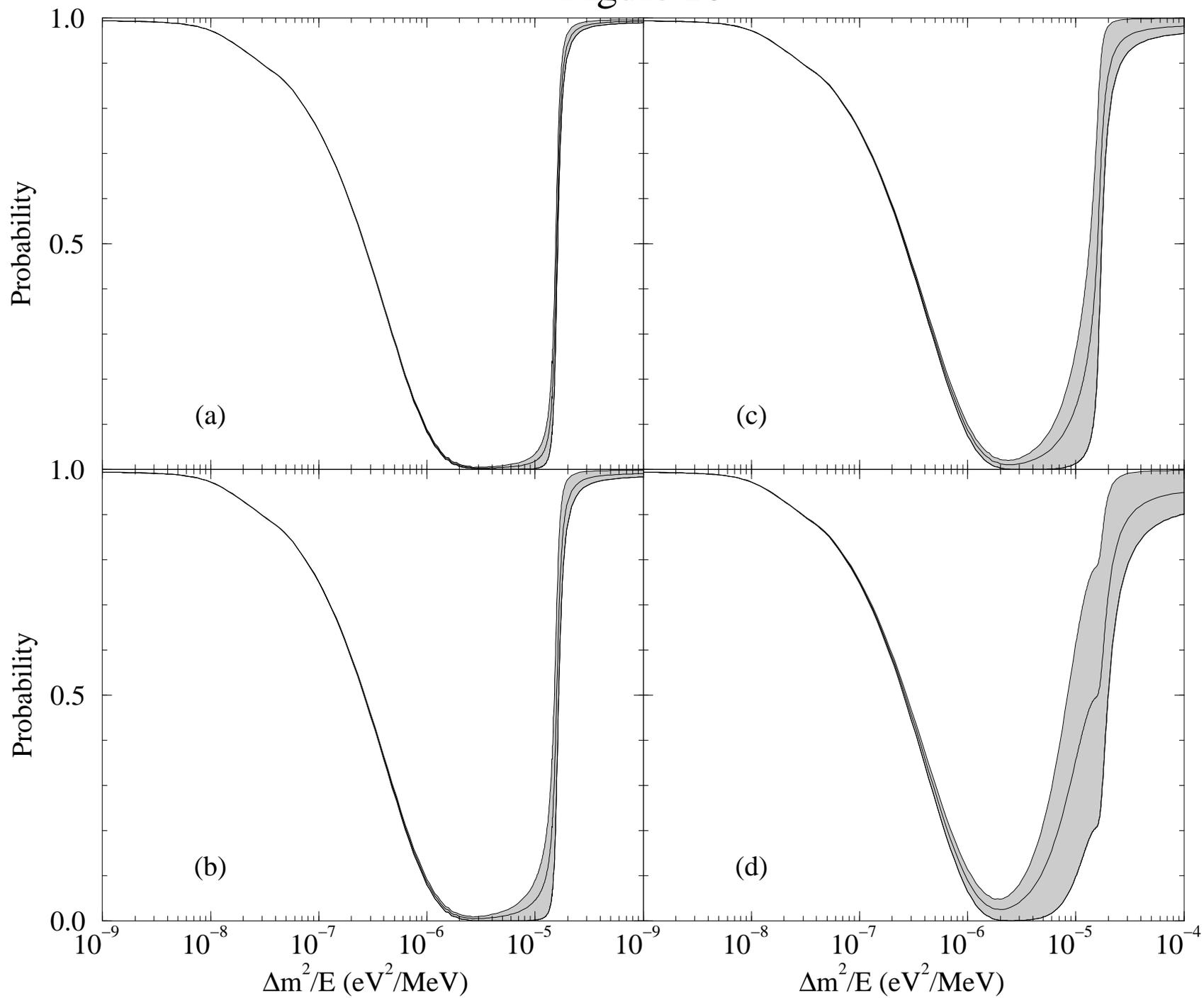
{\bf Figure 10.} Mean survival probability plus and minus $\sigma$
for the Bahcall-Pinsonneault density profile with $\sin^2 2\theta =
0.01$.  (a)-(d) show $\langle \beta \rangle_{\hbox{\scriptsize rms}}$
equal to 0.01, 0.02, 0.04, and 0.08, respectively.
\bigskip

\noindent
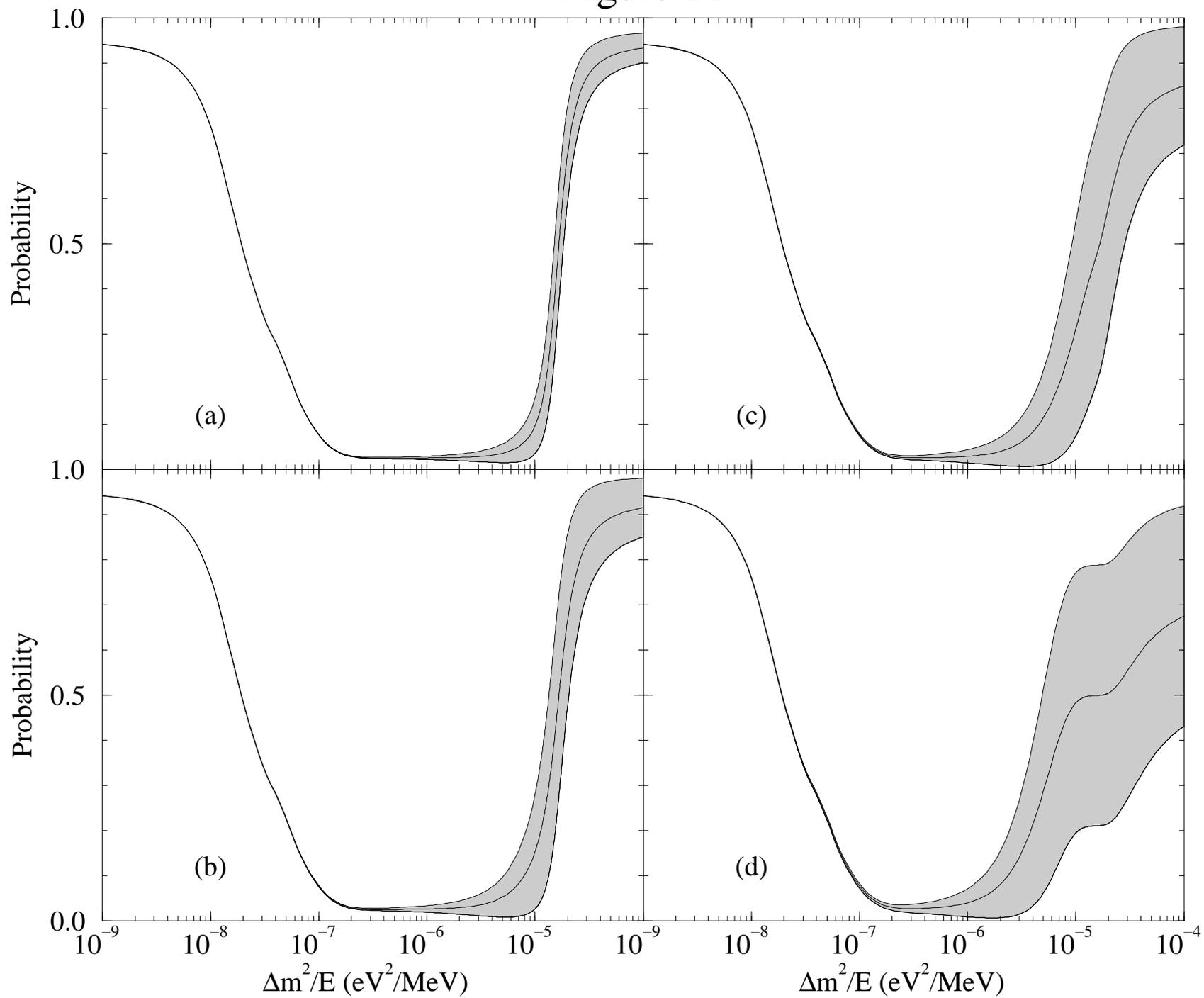
{\bf Figure 11.} As Figure 10, but with $\sin^2 2\theta = 0.1$.

\noindent
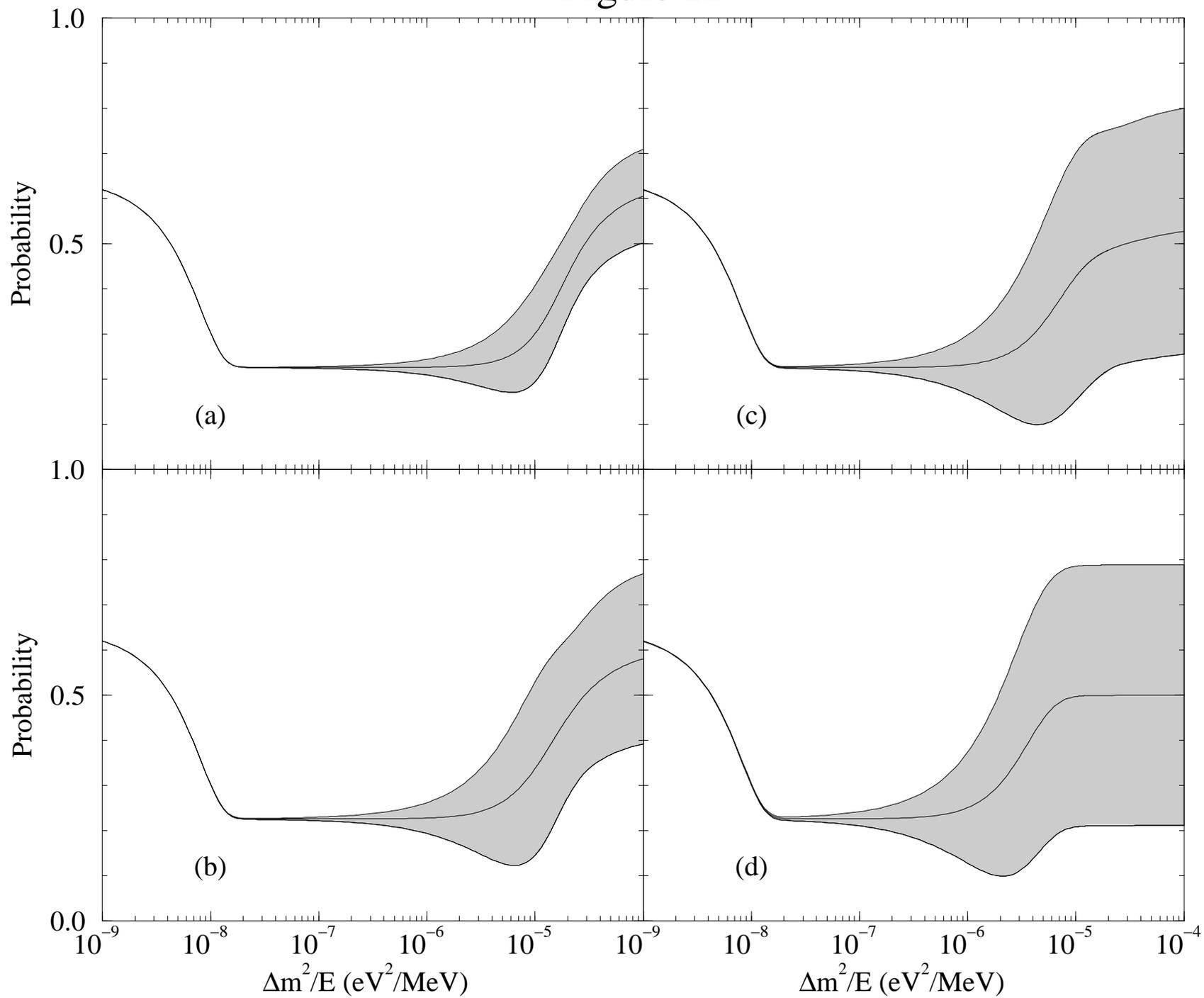
{\bf Figure 12.} As Figure 10, but with $\sin^2 2\theta = 0.7$.

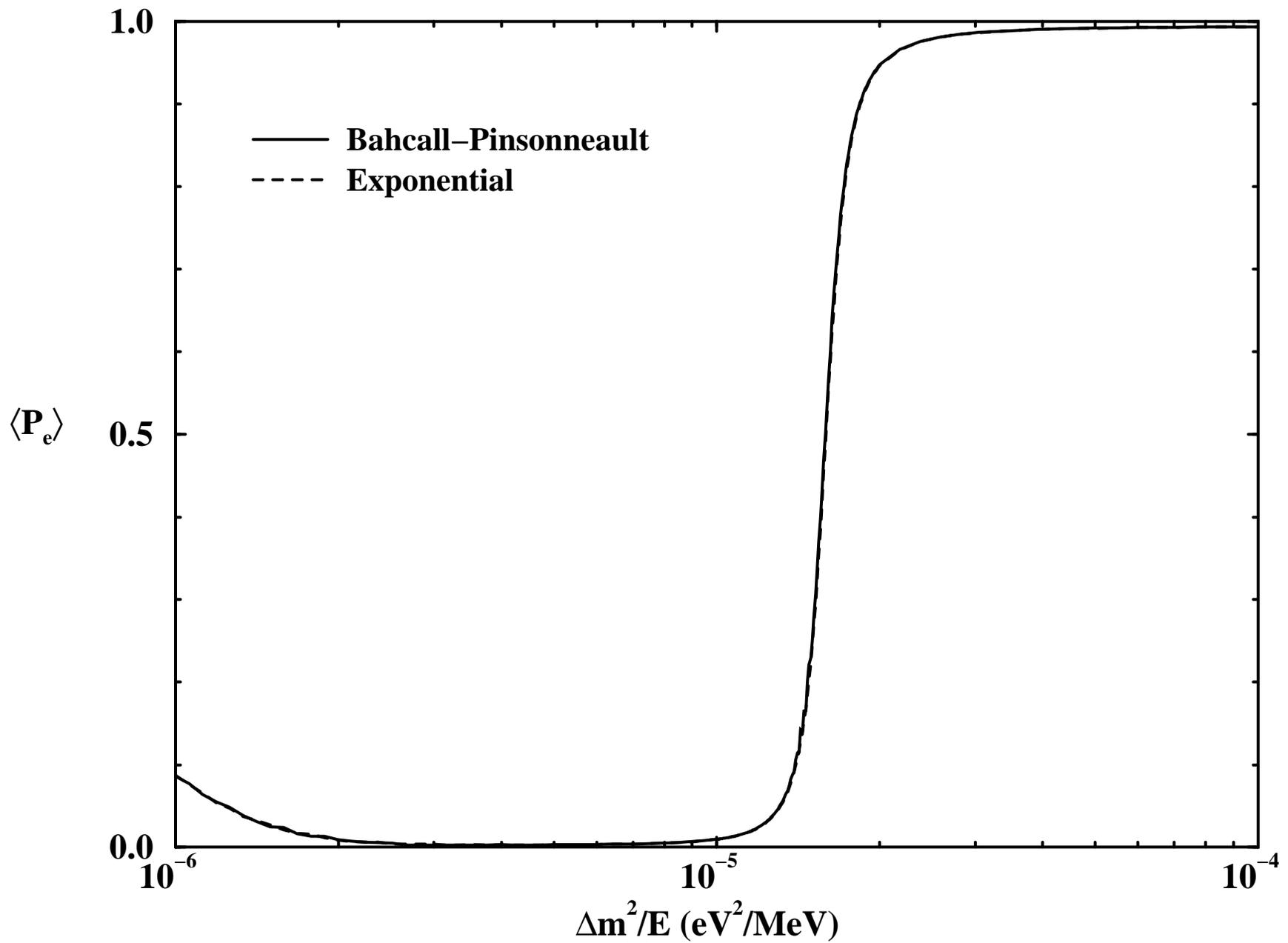
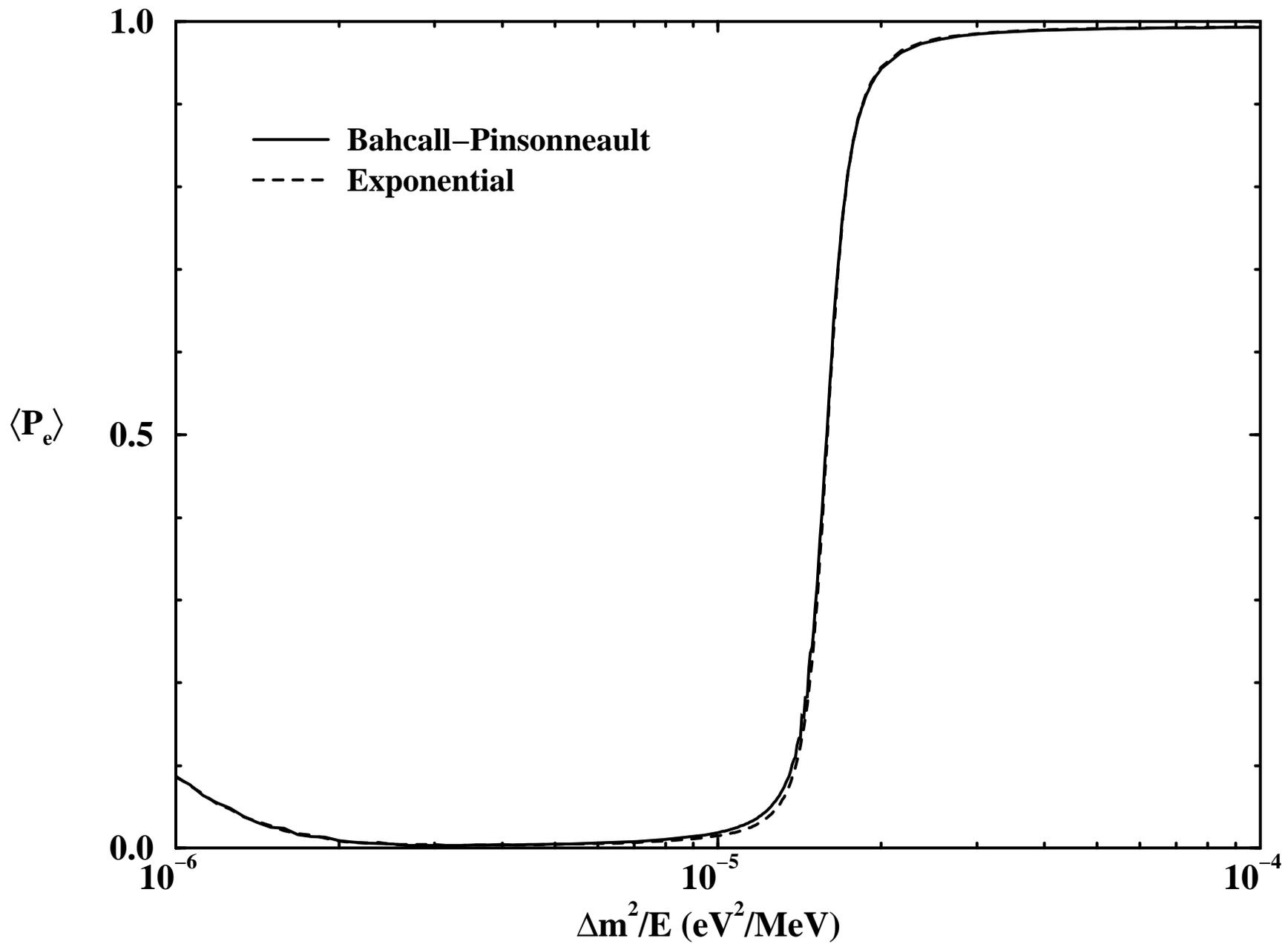
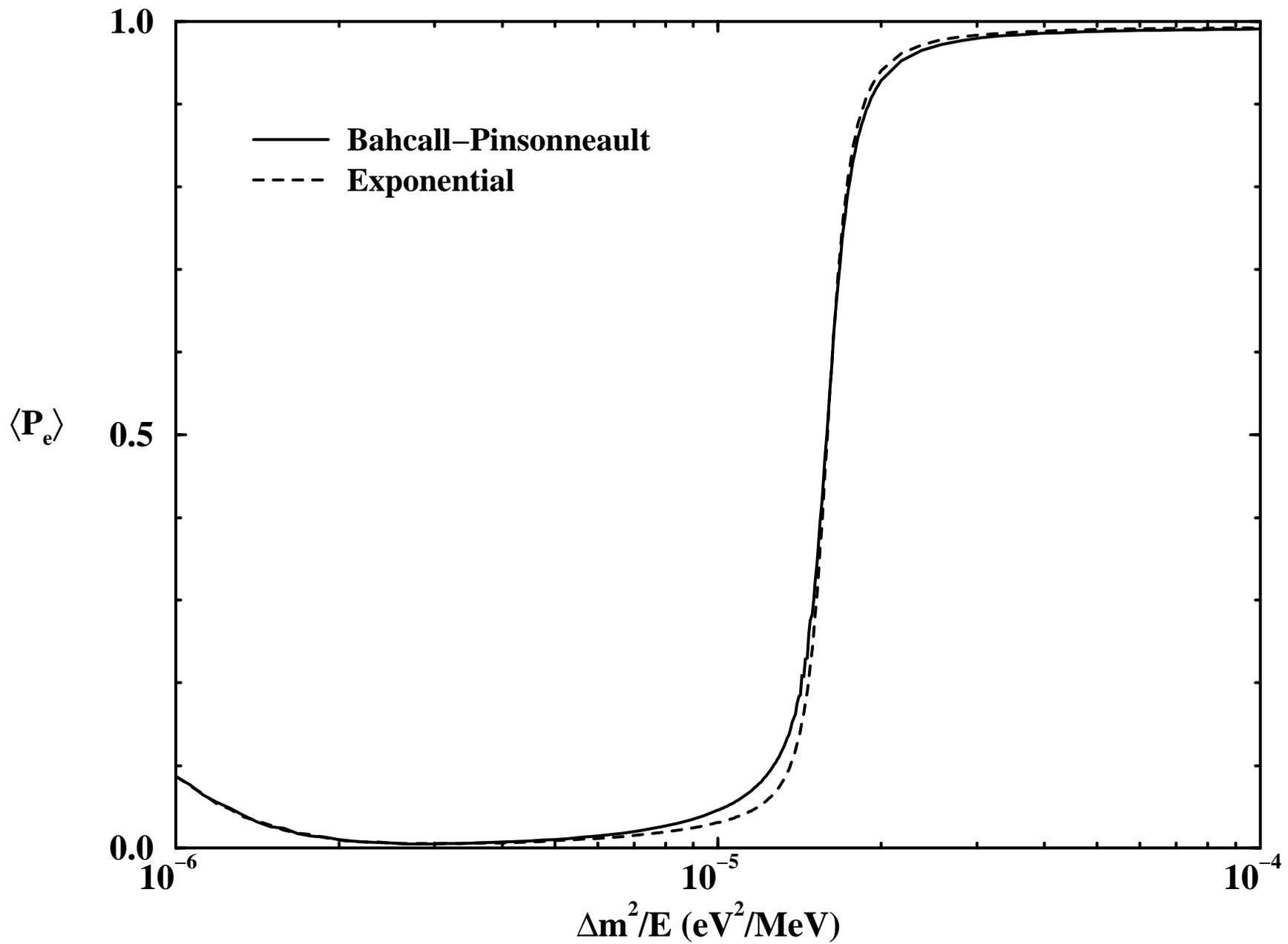
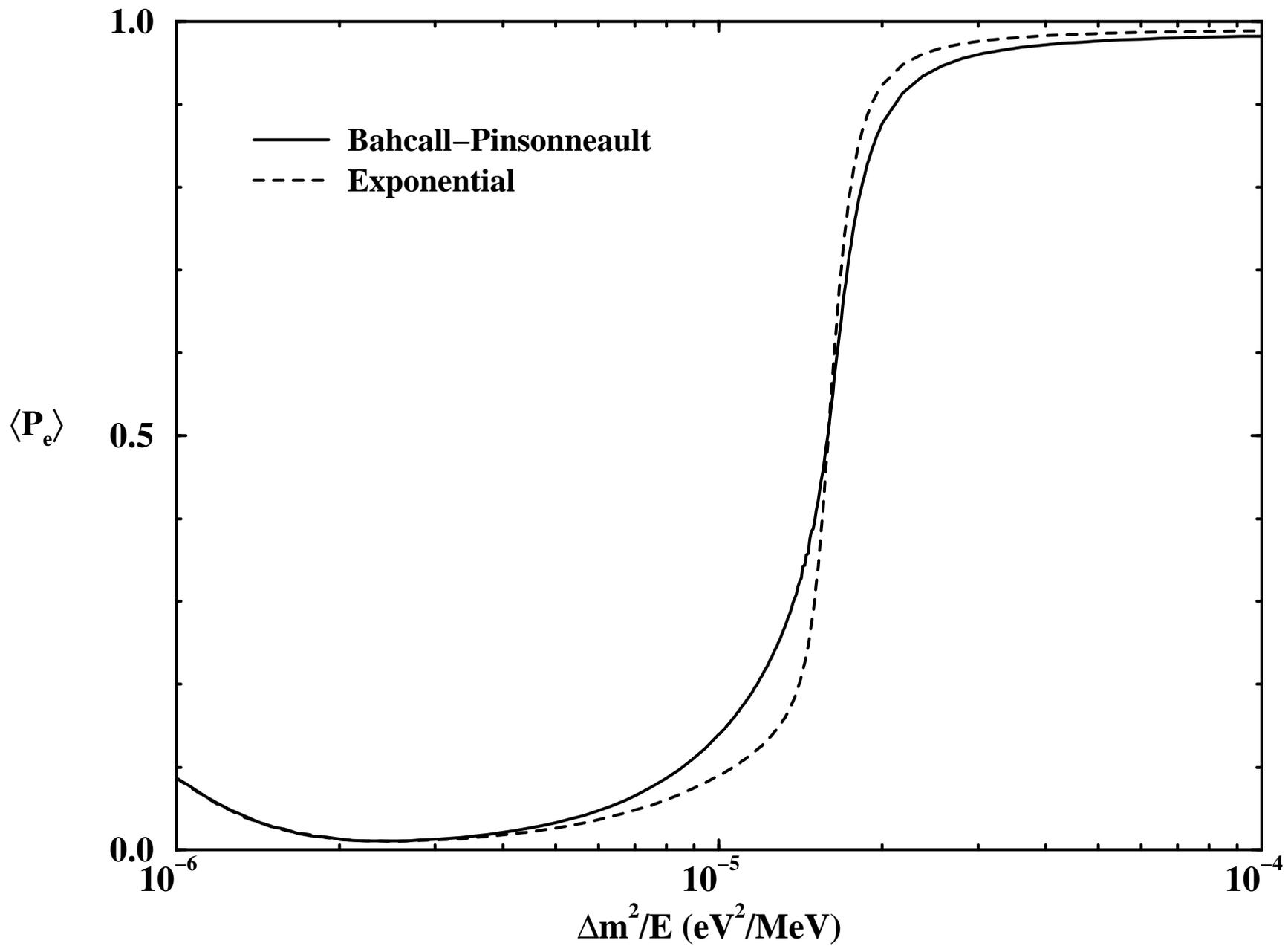
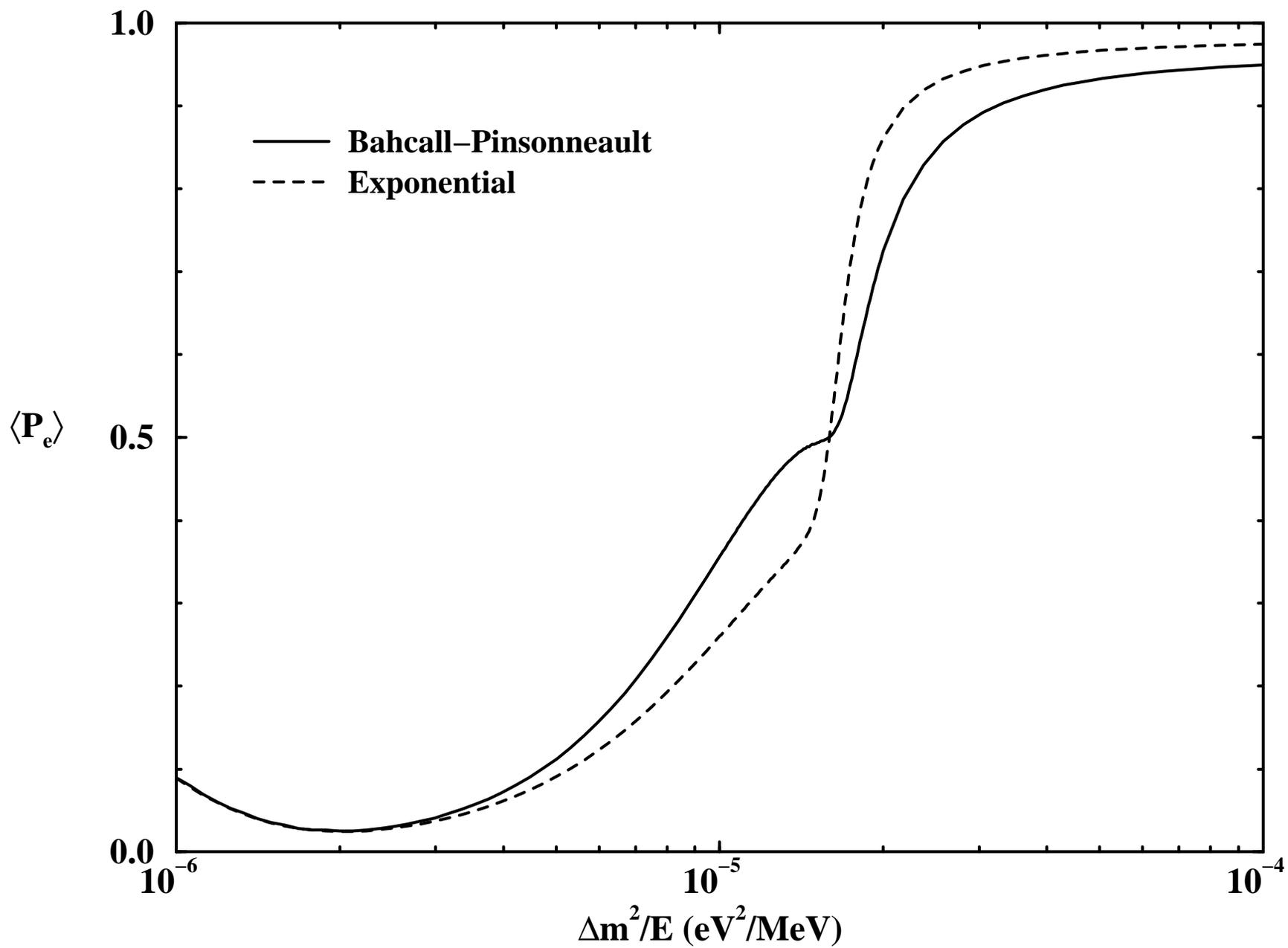
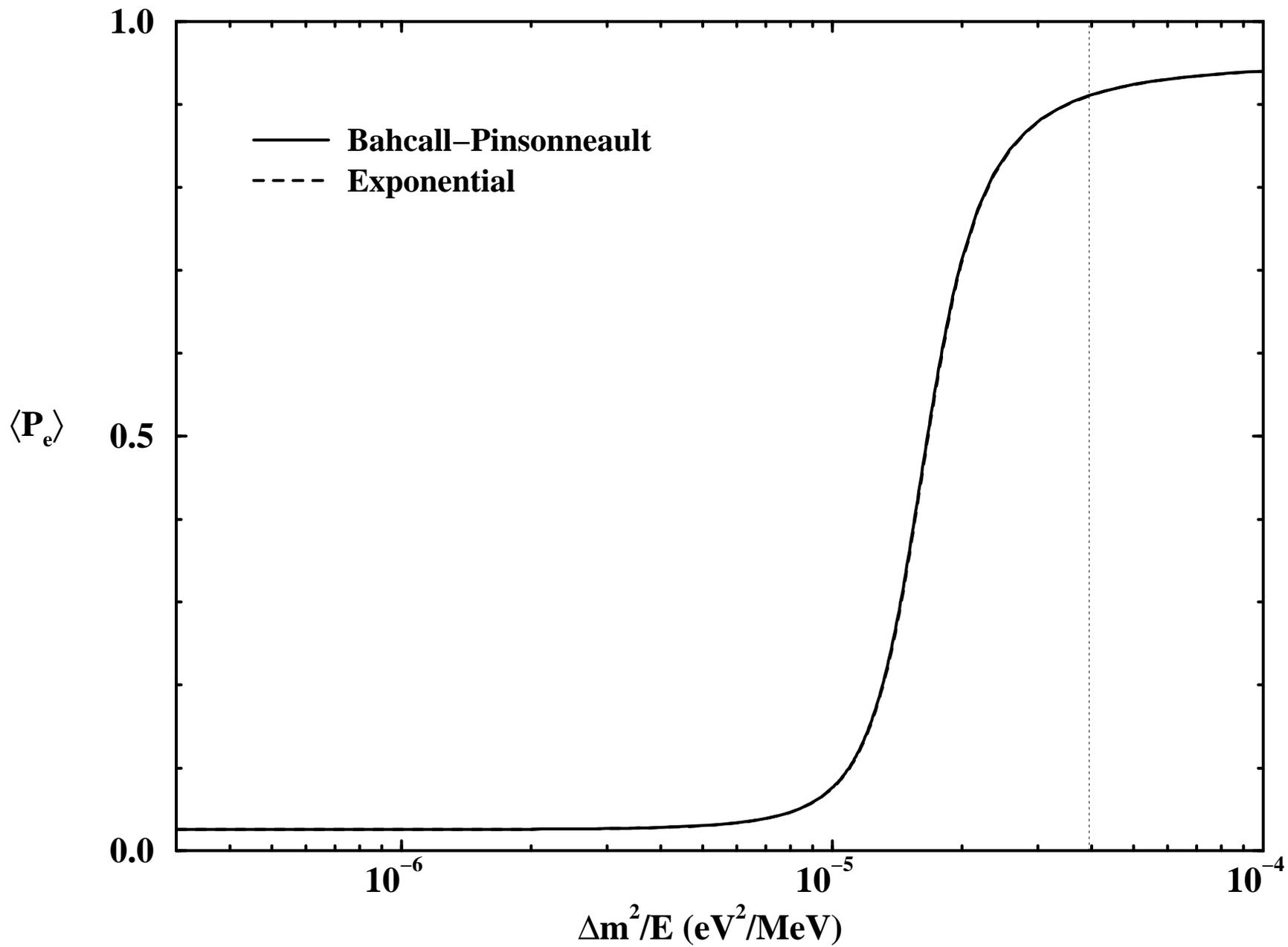
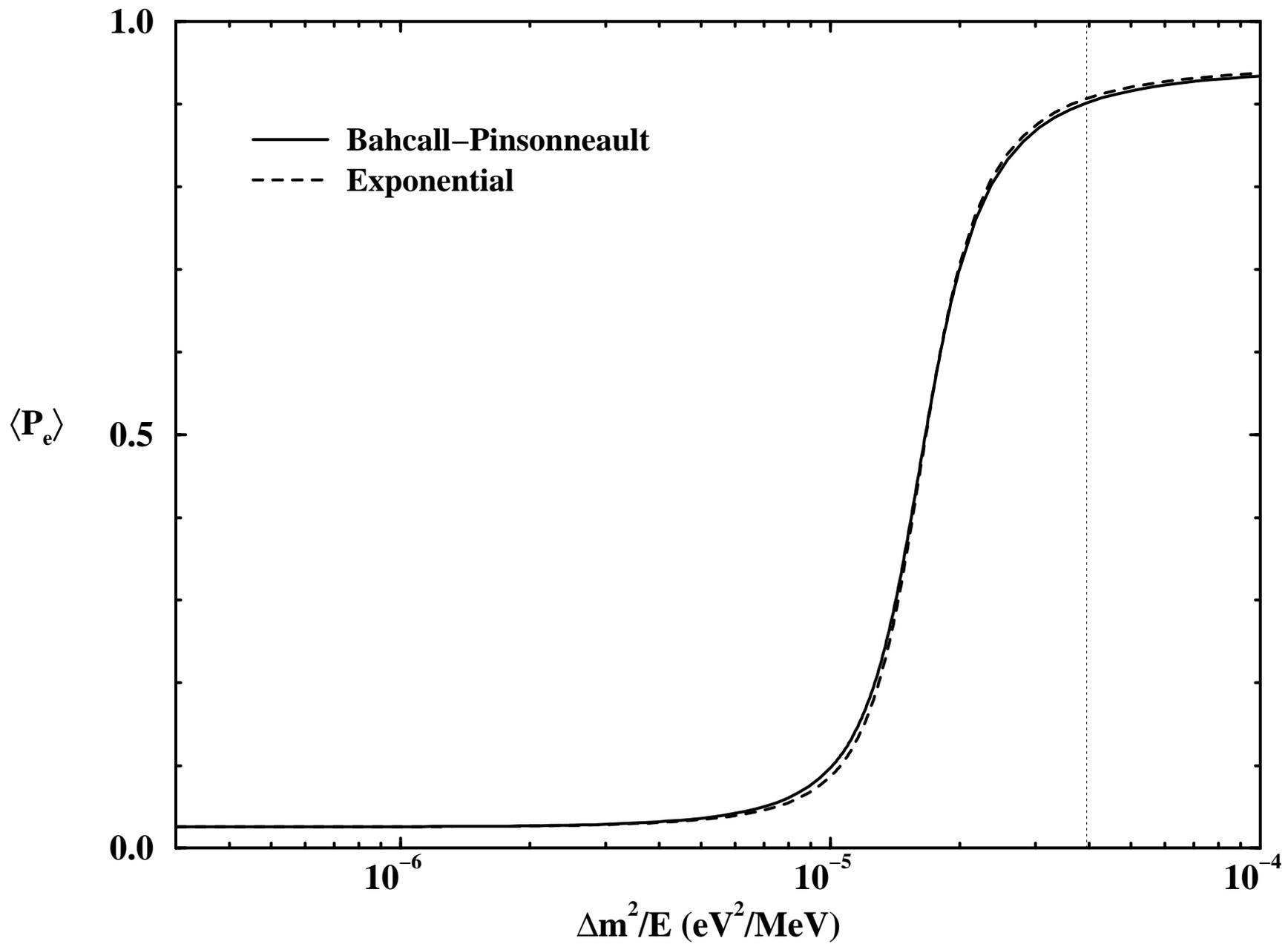
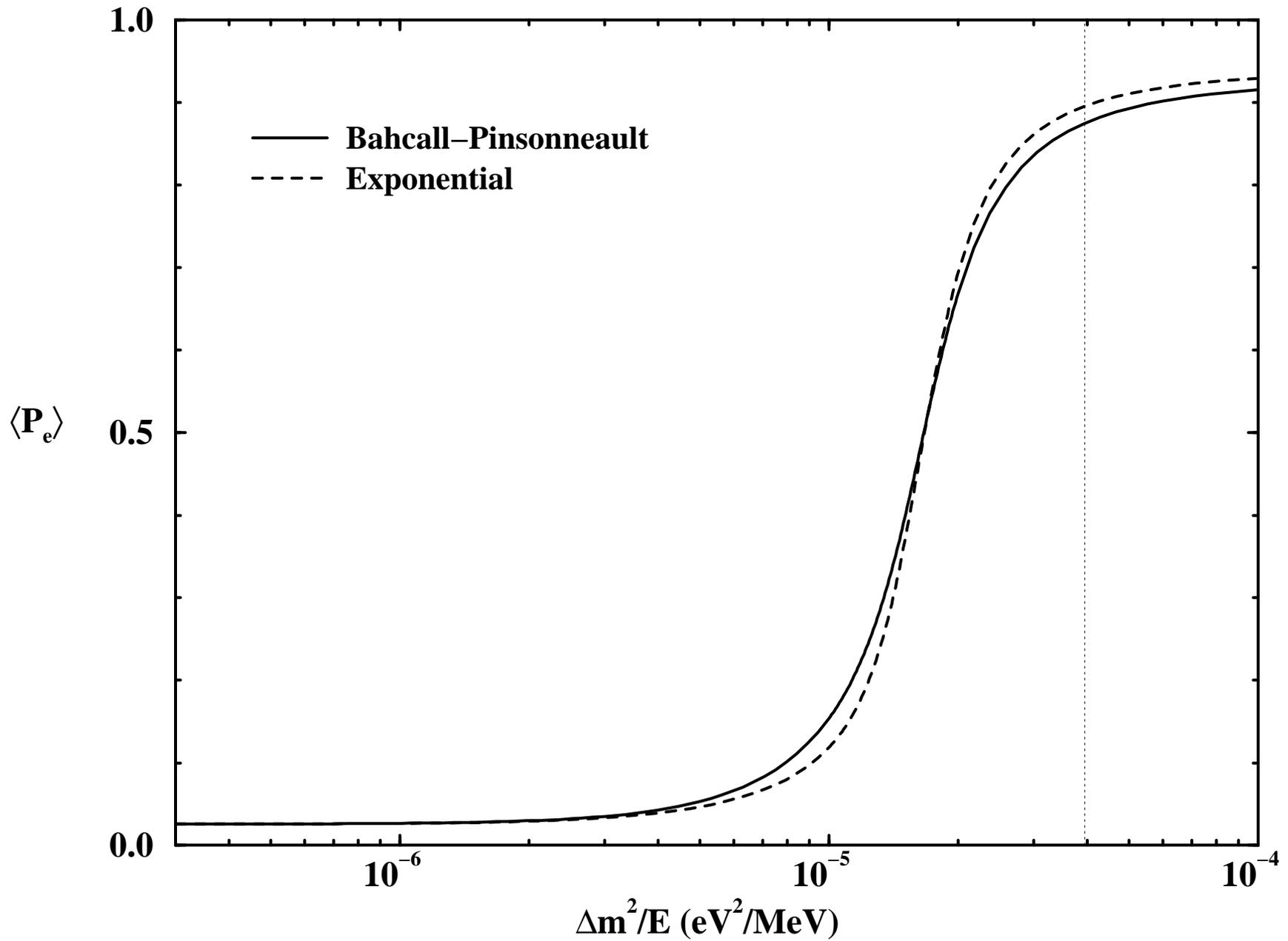
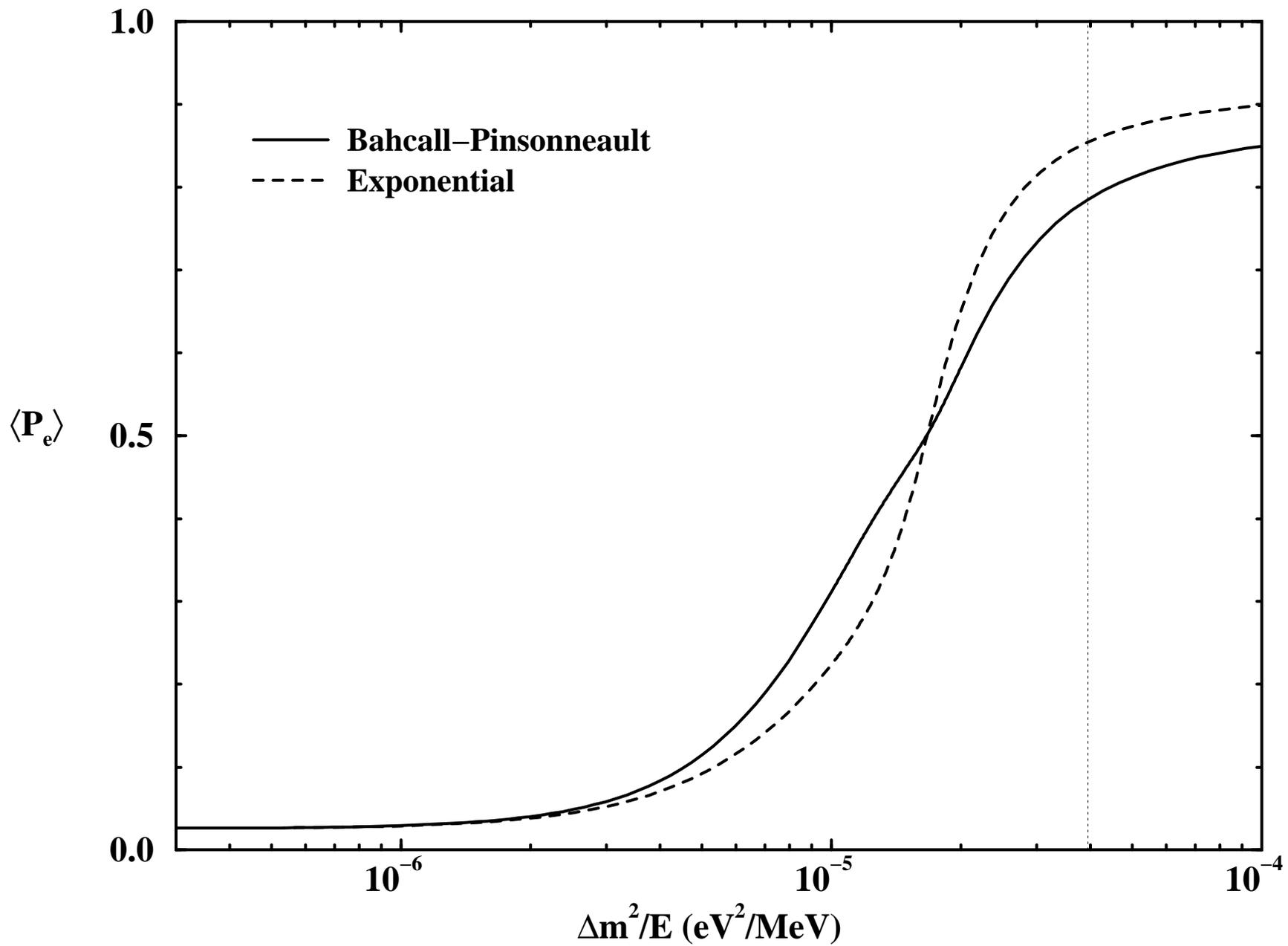
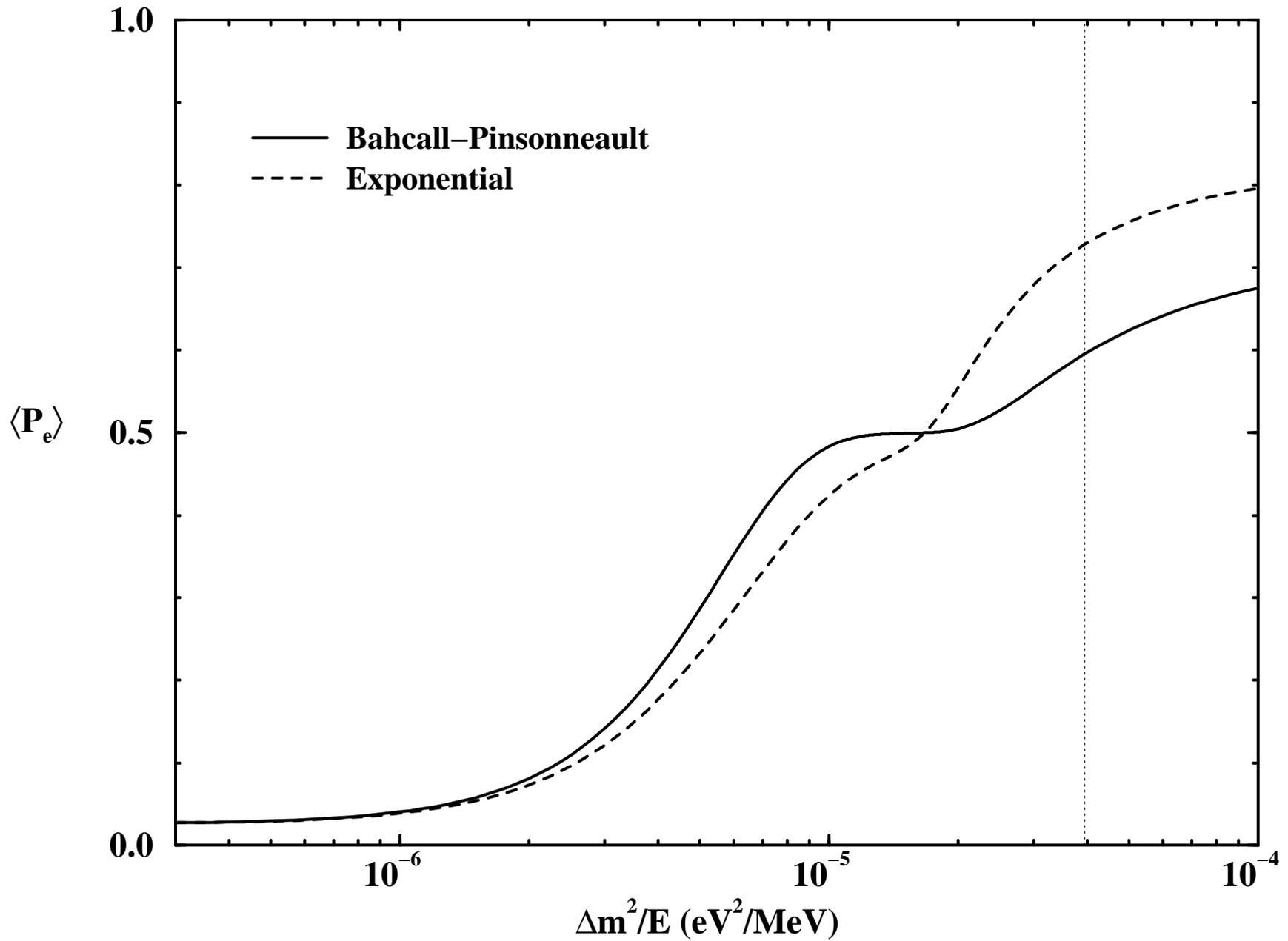
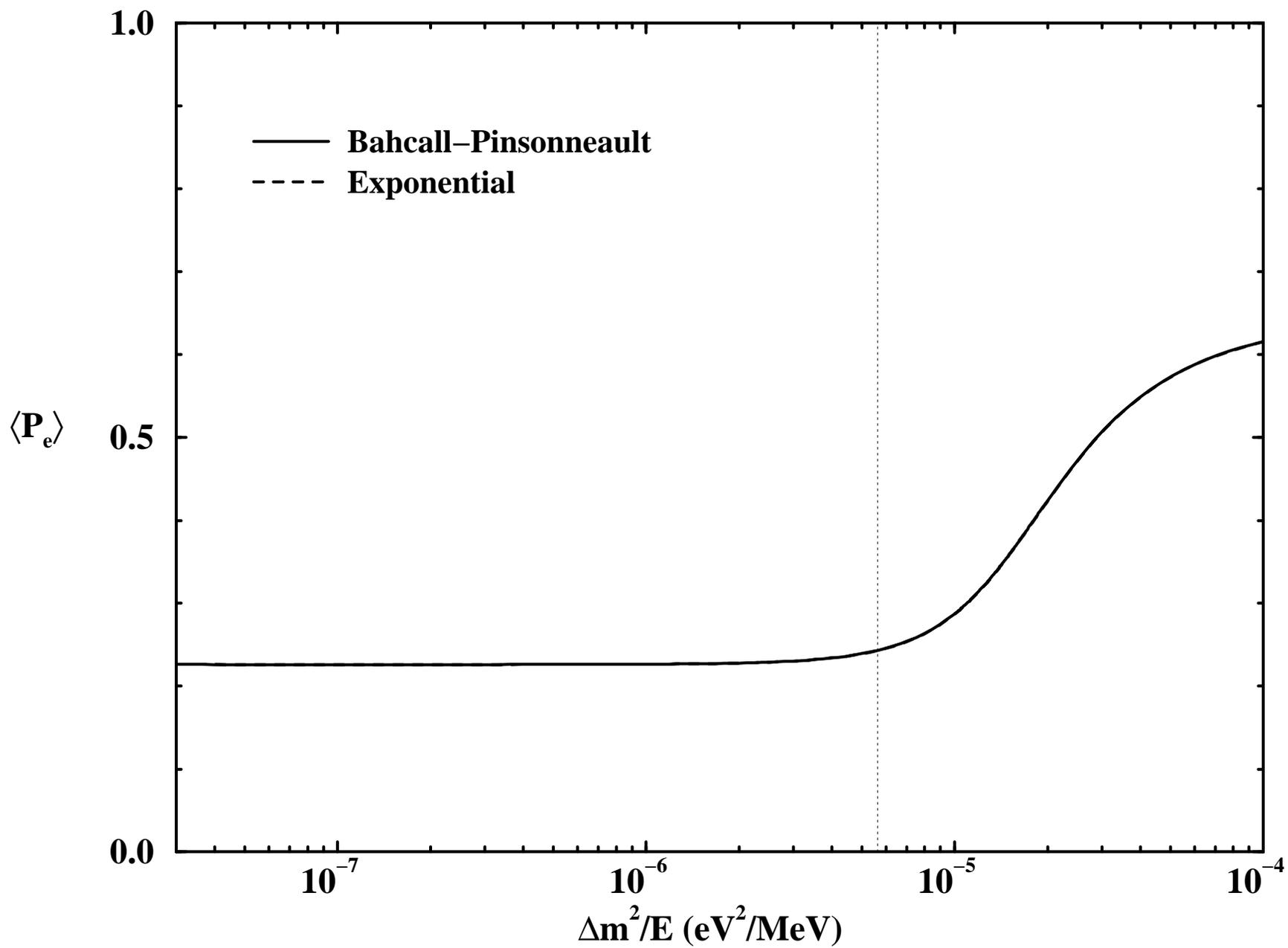
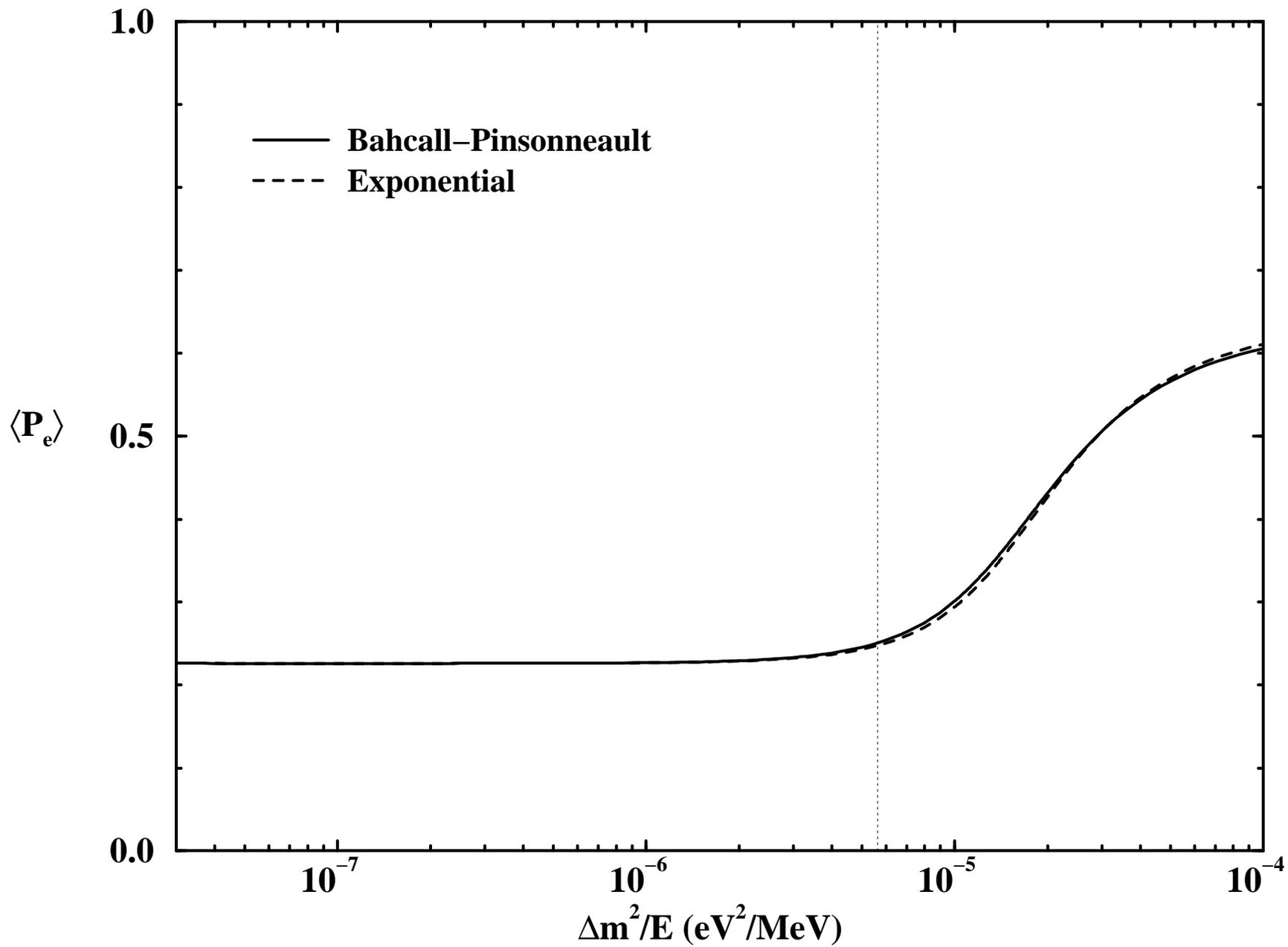
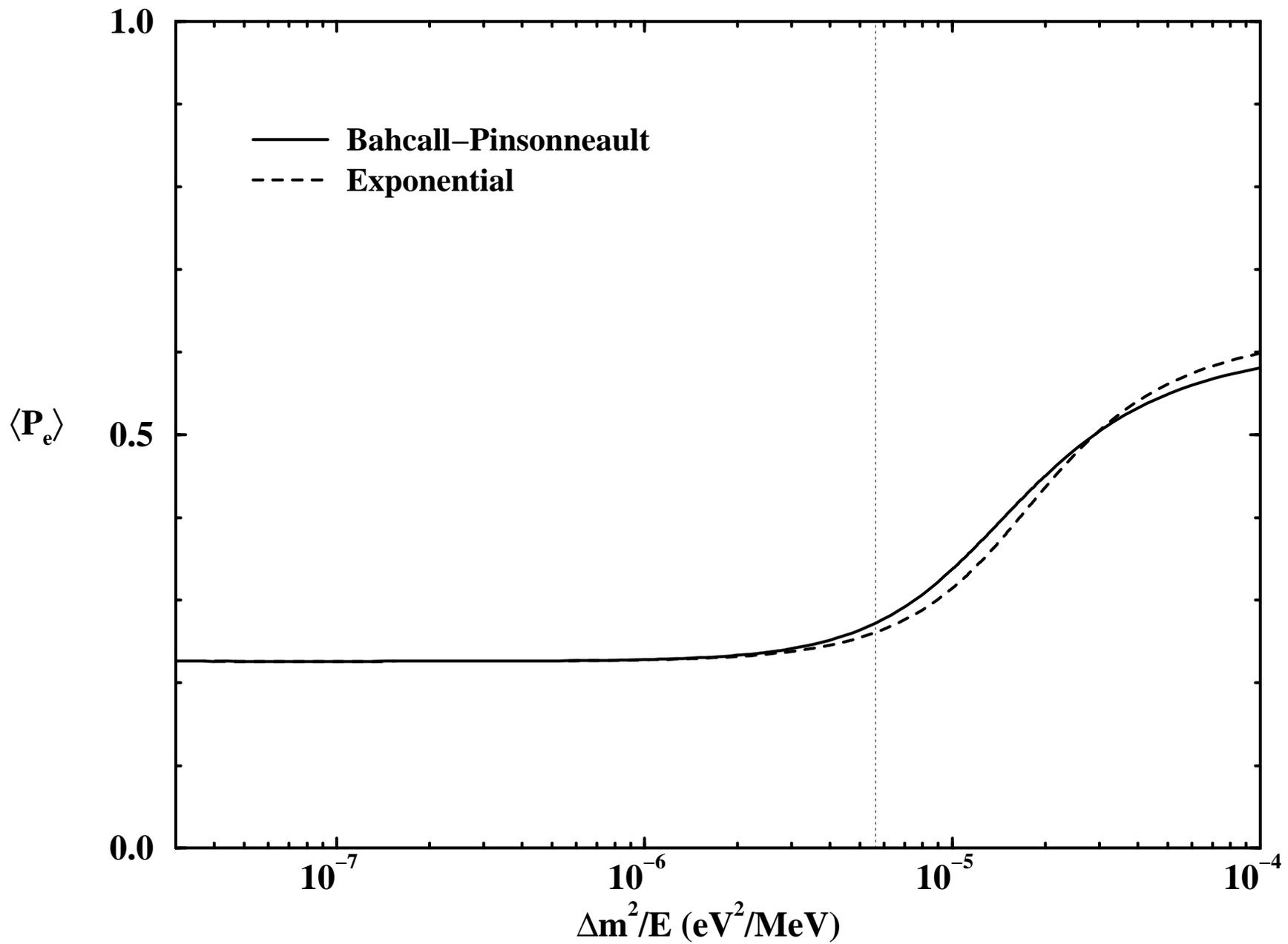
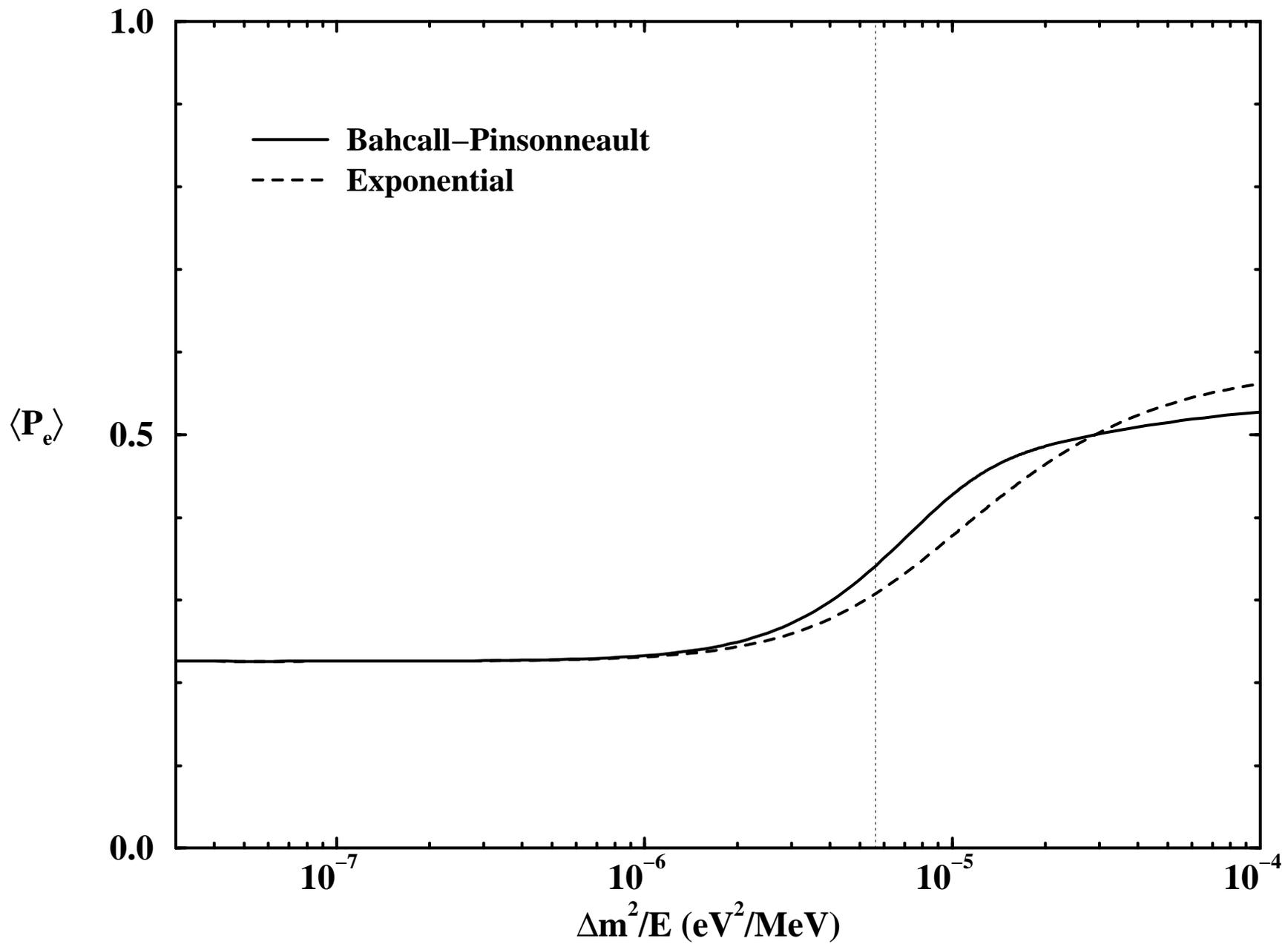
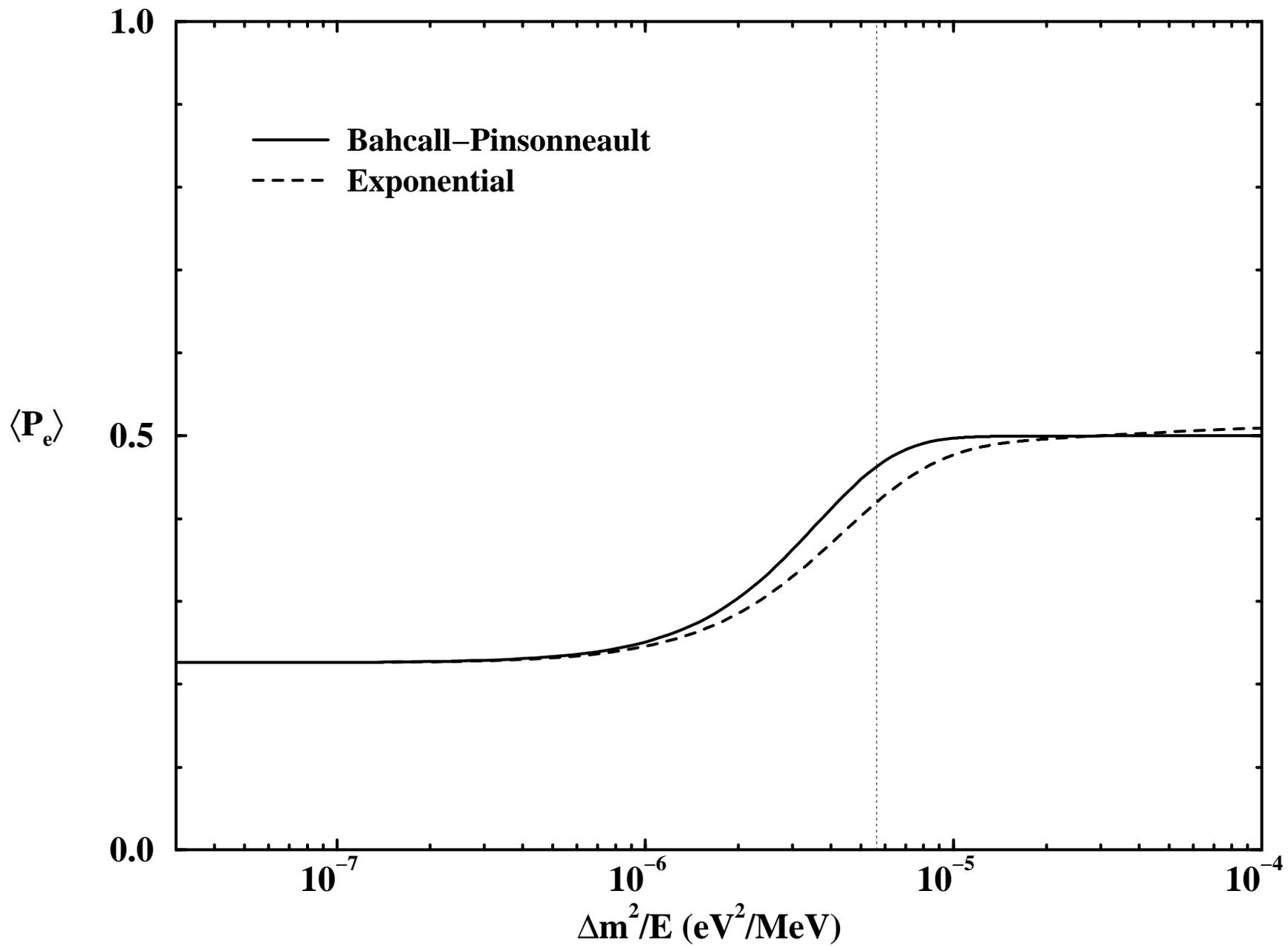
\end{document}